\documentclass[a4paper,fleqn,usenatbib,useAMS]{mnras}

\usepackage{graphicx}
\usepackage{amssymb}
\usepackage{subfigure}
\usepackage{captcont}
\usepackage{multirow}
\usepackage{longtable}
\usepackage{comment}
\usepackage{array}
\usepackage{url}
\usepackage{lscape}
\usepackage{booktabs,multirow}
\usepackage{times} 

\usepackage{aas_macros} 

\usepackage[T1]{fontenc}
\usepackage{aecompl}
\usepackage{newtxtext,newtxmath}
\usepackage[usenames]{color}

\newcommand{\cd}{d$^{-1}$}

\title[WET observations of the roAp star J1940]{Whole Earth Telescope discovery of a strongly distorted quadrupole pulsation in the largest amplitude rapidly oscillating Ap star}
\author[D. L. Holdsworth et al.]{Daniel L. Holdsworth,$^{1,2}$\thanks{E-mail:dlholdsworth@uclan.ac.uk}
D.W. Kurtz,$^{1}$
H. Saio,$^{3}$
J.L. Provencal,$^{4}$
B. Letarte,$^{5}$
\newauthor
R. Sefako,$^{6}$
V. Petit,$^{7,4}$
B. Smalley,$^{2}$
H. Thomsen$^{7}$
and C. L. Fletcher$^{7}$\\
$^{1}$ Jeremiah Horrocks Institute, University of Central Lancashire, Preston PR1 2HE, UK\\
$^{2}$ Astrophysics Group, Keele University, Staffordshire ST5 5BG, UK\\
$^{3}$ Astronomical Institute, School of Science, Tohoku University, Sendai 980-8578, Japan\\
$^{4}$ Department of Physics and Astronomy, University of Delaware, Newark DE 19716, USA\\
$^{5}$ Department of Physics, North-West University, Mafikeng Campus, Private Bag X2046, Mmabatho 2745, South Africa\\
$^{6}$ South African Astronomical Observatory, PO Box 9, Observatory, Cape Town 7935, South Africa\\
$^{7}$ Florida Institute of Technology, 150 W. University Blvd., Melbourne, FL 32907, USA
}
\begin{document}

\date{\today}

\pagerange{\pageref{firstpage}--\pageref{lastpage}} \pubyear{2017} 

\maketitle

\label{firstpage}

\begin{abstract}
We present a new analysis of the rapidly oscillating Ap (roAp) star, 2MASS\,J$19400781-4420093$ (J1940; $V=13.1$). The star was discovered using SuperWASP broadband photometry to have a frequency of 176.39\,\cd\, (2041.55\,$\muup$Hz; $P = 8.2$\,min; \citealt{holdsworth14a}) and is shown here to have a peak-to-peak amplitude of 34\,mmag. J1940 has been observed during three seasons at the South African Astronomical Observatory, and has been the target of a Whole Earth Telescope campaign. The observations reveal that J1940 pulsates in a distorted quadrupole mode with unusual pulsational phase variations. A higher signal-to-noise ratio spectrum has been obtained since J1940's first announcement, which allows us to classify the star as A7\,Vp\,Eu(Cr). The observing campaigns presented here reveal no pulsations other than the initially detected frequency. We model the pulsation in J1940 and conclude that the pulsation is distorted by a magnetic field of strength $1.5$\,kG. A difference in the times of rotational maximum light and pulsation maximum suggests a significant offset between the spots and pulsation axis, as can be seen in roAp stars.
\end{abstract}


\section{Introduction}
\label{sec:intro}

In the region of the Hertzsprung-Russell (HR) diagram where the classical instability strip crosses the main-sequence, a plethora of stellar variability is found. This intersection occurs at the temperature range where the A and F stars lie. These stars exhibit an array of spectral abnormalities, from the metal deficient $\lambda$\,Bo\"otis stars to the chemically peculiar Ap stars, and show a variety of variability frequencies, from the low-frequency $\gamma$\,Doradus pulsators to the higher frequency $\delta$\,Scuti stars, through to the rapidly oscillating Ap (roAp) stars.

\begin{figure}
\includegraphics[width=\linewidth]{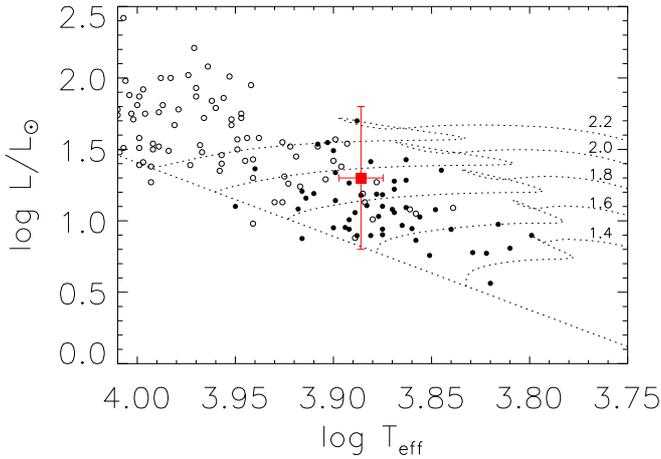}
\caption{The positions of the roAp stars for which GAIA or Hipparcos parallaxes are available (filled circles). We include the non-oscillating Ap stars for comparison (open circles) for which there is reliable data, and non-detections, in the literature. The subject of this study, J1940, is shown by the red square, with its position determined in sections\,\ref{sec:spec} and \ref{sec:modelling}. The zero-age main-sequence and the evolutionary tracks are from \citet{bertelli08}. Note, we do not plot the error bars of the other stars for clarity; these are typically $\pm0.02$ and $\pm0.3$ in $\log T_{\rm eff}$ and $\log L/L_{\odot}$, respectively.}
\label{fig:HRD}
\end{figure}

The roAp stars are a rare subclass of the chemically peculiar, magnetic, Ap stars which show pulsations in the range of $6-23$\,min with amplitudes up to 18\,mmag peak-to-peak in Johnson $B$ \citep{holdsworth15}. Their luminosities range from the zero-age main-sequence to beyond terminal-age main-sequence. Since their discovery by \citet{kurtz82}, only 61 of these objects have been identified (see \citealt{smalley15} for a catalogue). Fig.\,\ref{fig:HRD} shows the position of the roAp stars on the HR diagram for which GAIA or Hipparcos parallaxes are available.

The pulsations in roAp stars are high-overtone pressure modes (p\,modes) thought to be driven by the $\kappa$-mechanism acting in the H\,{\sc{i}} ionisation zone \citep{balmforth01}. However, \citet{cunha13} have shown that turbulent pressure in the convective zone may excite some of the modes seen in a selection of roAp stars.

The roAp stars are unique amongst pulsating stars as their pulsation axis is inclined to the rotation axis, and closely aligned with the magnetic one, leading to the oblique pulsator model \citep[OPM; ][]{kurtz82,ss85a,ss85b,dg85,shibahashi93,ts94,ts95,bigot02,bigot11}. As such, the pulsation modes can be viewed from varying aspects over the rotation cycle of the star, leading to a modulated pulsation amplitude which gives constraints on the pulsation geometry that are not available for any other type of pulsating star (other than the Sun, which is uniquely resolved). 

The Ap stars show very strong mean magnetic field moduli, of the order a few kG to 34\,kG \citep{babcock60,mathys17}. Such a strong magnetic field is able to suppress convection, thus providing a stable environment in which radiative levitation can occur. This mechanism leads to a stratified atmosphere with significant surface inhomogeneities, often, in the case of the roAp stars \citep{ryabchikova04}, consisting of singly and doubly ionised rare earth elements. In these inhomogeneities, or spots, elements such as La, Ce, Pr, Nd, Sm, Eu, Gd, Tb, Dy and Ho, may be overabundant by up to a million times the solar value. In the presence of the magnetic field, these spots are very long lasting (decades in many known cases) allowing for an accurate determination of the rotation period of the star. These spots also cause spectral line strength variations over the rotation period of the star as different patches of elements drift in and out of view \citep[e.g.][]{lueftinger10}. Because of the complex atmospheres of the Ap stars, the roAp stars provide the best laboratory to study the interactions between pulsations, rotation, and chemical abundances in the presence of magnetic fields.

With the desire to expand the number of observed roAp stars, many photometric campaigns have targeted known Ap stars in the search for oscillations \citep[e.g.][]{martinez91,martinez94,handler99,dorokhova05,paunzen15,joshi16}. Later, with the advent of high-resolution and high-precision spectroscopy, Ap stars were monitored for line profile variations caused by pulsational velocity shifts \citep[e.g.][]{savanov99,koch01,hatzes04,mkr08,elkin10,elkin11,kochukhov13}. Finally, the use of the all-sky SuperWASP (Wide Angle Search for Planets) ground-based photometric survey led to the discovery of 11 roAp stars \citep{holdsworth14a,holdsworth15}. The use of such surveys removes previous biases, such as targeting cool Ap stars, when searching for these rare pulsators, allowing a broader parameter space to be explored.

With the launch of the {\it Kepler} space telescope, the ability to reach $\muup$mag precision enabled the detection of four roAp stars with pulsation amplitudes below the ground-based detection limit: KIC\,8677585 \citep{balona11a}; KIC\,10483436 \citep{balona11b}; KIC\,10195926 \citep{kurtz11}; and KIC\,4768731 \citep{smalley15}. {\it Kepler} observations also allowed for the detailed analysis of one roAp star, KIC\,7582608, identified in the SuperWASP survey with an amplitude of 1.45\,mmag \citep{holdsworth14b}, albeit in the super-Nyquist regime \citep{murphy13}.

Ground-based projects in the search for transiting exoplanets produce vast amounts of data on millions of stars (e.g. WASP, \citealt{pollacco06}; HATnet, \citealt{bakos04}; ASAS, \citealt{pojmanski97}; OGLE, \citealt{udalski92}; KELT, \citealt{pepper07}). These data can provide an excellent source of information on many thousands of variable stars. Indeed, many of these projects have been employed for that purpose \citep[e.g.][]{pepper08,hartman11,ulaczyk13,holdsworth14a,smalley17,holdsworth17a}. The ability of these surveys to achieve mmag precision provides an extensive all-sky database in which to search for low-amplitude stellar variability, which can then be observed at much higher precision by space-based missions such as {\it K2} \citep{howell14} and the upcoming Transiting Exoplanet Survey Satellite \citep[{\it TESS};][]{ricker15}. 

In the case of {\it K2} follow-up observations, two known roAp stars have been observed. HD\,24355 was discovered by \citet{holdsworth14a} and subsequently observed in the short cadence (SC) mode during Campaign 4. Analysis of those observations led to the conclusion that HD\,24355 is a highly distorted quadrupole pulsator with very unusual pulsational phase variations over the rotation period \citep{holdsworth16}. The other star, HD\,177765, was observed during Campaign 7 in the long cadence (LC) mode. That star was discovered to be a roAp star through spectroscopic observations \citep{alentiev12}. Previous ground-based photometric observations of HD\,177765 failed to detect the pulsation; however, as spectroscopy has the ability to detect smaller amplitudes, HD\,177765 was shown to be a roAp star. The {\it K2} space-based observations do, however, have the precision to detect low amplitude pulsations. \citet{holdsworth16b} used the {\it K2} data to confirm the known pulsation with photometry, and to identify two further, low-amplitude, pulsations. It is only with near-continuous, highly precise observations that such a result could be obtained.

In an attempt to reduce, or ideally remove, aliasing that affects the analysis of ground-based time-series observations, the Whole Earth Telescope (WET) was established \citep{nather90}. The WET is an international collaboration that aims to achieve 24\,hr coverage of pulsating stars through a network of telescopes positioned at different longitudes around the globe. The data are collected at several observing sites and fed back to headquarters where uniform data reduction is conducted. This strategy aims to provide continuous data of a star, thus removing alias ambiguity from frequency analysis. The WET is a good method to gain a high duty-cycle of a star without using space-based telescopes. 

The subject of this paper, J1940 ($\alpha$:\,19:40:07.81, $\delta$:\,$-44$:20:09.3; 2MASS\,J19400781-4420093), is a relatively faint ($V=13.1$) roAp star discovered by \citet{holdsworth14a} through a survey of A stars in the SuperWASP archive. Their data show a pulsation at 176.39\,\cd\, ($2041.55\,\muup$Hz; $P=8.2$\,min) with an amplitude of 4.16\,mmag in the WASP broadband filter. The pulsation amplitude of roAp stars depends strongly on the filter used for the observations \citep{medupe98}. As such, when considering the filter differences, J1940 is the largest amplitude roAp star observed to date. Such a significant pulsation made this star a prime target for WET observations. In this paper we present an in-depth discussion of the SuperWASP discovery data, alongside further ground-based observations, and the result of the WET observations. We also provide a more accurate spectral classification of this star than previously published.


\section{Spectral Classification}
\label{sec:spec}

We have obtained two low-resolution spectra of J1940 with the Robert Stobie Spectrograph \citep[RSS; ][]{kobulnicky03} mounted on the Southern African Large Telescope (SALT). The first, published by \citet{holdsworth14a}, was taken on 2012 November 3 with an exposure time of 759\,s leading to a signal-to-noise ratio (S/N) of $\sim$30 and a two pixel resolution of 0.37\,\AA. The second spectrum was obtained on 2015 May 21 with an exposure time of 2581\,s leading to a S/N of $\sim$80 and a two pixel resolution of 0.73\,\AA. Both spectra were taken with similar instrumental set ups, using the PG2300 grating at an angle of $30.875^\circ$ and a camera angle of $61.75^\circ$. The only difference was the slit width: for the 2012 observation this was at 0.6\,arcsec, whereas in 2015 it was 1\,arcsec. Table\,\ref{tab:spec} shows a log of the observations.

\begin{table}
  \caption{Details for spectral observations of J1940. The S/N was determined using the  {\sc{der\_snr}} code of \citet{stoehr08}. The rotation phase has been calculated from equation\,(\ref{equ:rot}), as shown in the text.}
   \centering
  \label{tab:spec}
  \begin{tabular}{ccccc}
    \hline
    BJD-245\,0000.0 & Exposure time & S/N & Resolution & Rotation\\
                              &   (s)         &     &  (\AA)          & phase   \\
    \hline
    6235.2708 & 759   & 30 & 0.37  & $0.30\pm0.01$ \\
    7164.4569 & 2851 & 80 & 0.73  & $0.76\pm0.02$ \\
        \hline
    \end{tabular}
\end{table}
    
\begin{figure}
\includegraphics[width=\linewidth]{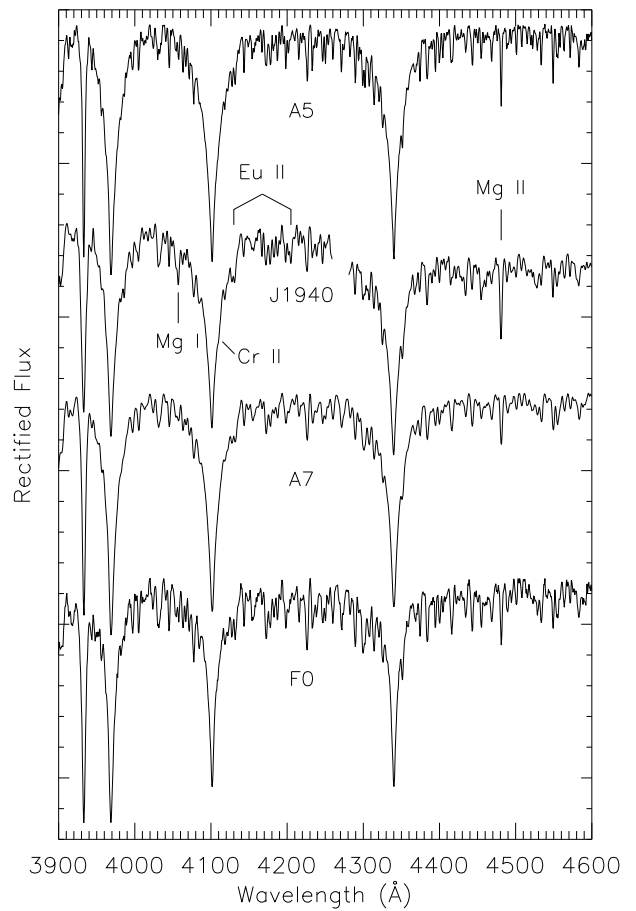}
\caption{Spectrum of J1940 (second) compared to MK standard stars of A5 type (top), A7 type (third) and F0 type (bottom). The target spectrum has been convolved with a Gaussian profile to the resolution of the MK standards (1.67\,\AA). The spectra have been offset for clarity.}
\label{fig:spec}
\end{figure}

We do not co-add the two spectra for this work. Although doing so increases the S/N (from $\sim$80 to $\sim$82), the two spectra were obtained at different rotation phases, and co-adding the spectra can dilute the signatures of chemical peculiarities (due to their spotty nature). The small increase in S/N does not affect our comparison to the standards, but diluting the chemical peculiarities will affect our conclusion. Therefore, we present the new spectrum in Fig.\,\ref{fig:spec}. Alongside the spectrum of J1940, we show spectra of three MK standard stars\footnote{Spectra of the standard stars were obtained from R.O. Gray's website: \url{http://stellar.phys.appstate.edu/Standards/}} (A5: HD\,23194; A7: HD\,23156; F0: HD\,23585). We determine that the A7 spectral class best fits the Balmer lines of J1940. There are deviations from the standard spectrum, however, as one would expect in an Ap star. Particularly, lines of Eu\,{\sc ii} at $4130$ and $4205$\,\AA\, are enhanced, with a tentative signature of Cr\,{\sc ii} at 4111\,\AA. This leads us to conclude that J1940 is an A7\,Vp\,Eu(Cr) star. Finally it should be noted the Mg\,{\sc i} feature at 4057\,\AA\, and the Mg\,{\sc ii} line at 4481\,\AA\, are significantly stronger than the MK standard. The Ca\,{\sc{ii}} K line is slightly weak in this star, which is common in Ap stars. The line is broader and shallower than that of the standard star.

Through fitting of the Balmer lines, at a fixed $\log g=4.0$ (cgs), we derive a temperature for J1940 of $7700\pm200$\,K. Fixing the $\log g$ at this value is reasonable, especially given the low resolution of the spectrum, and given its position in Fig.\,\ref{fig:HRD}. Where available, the $\delta c_1$ index from Str\"omgren photometry is used to provide an indication of the $\log g$ value of A stars. However, this is not reliable for the Ap stars due to heavy line blanketing which results in, on the whole, a negative value of $\delta c_1$ \citep[see e.g.][]{kurtz00,joshi16}. Hence we set the value of $\log g$ at $4.0$ when classifying the star.

\begin{figure}
\includegraphics[width=\linewidth]{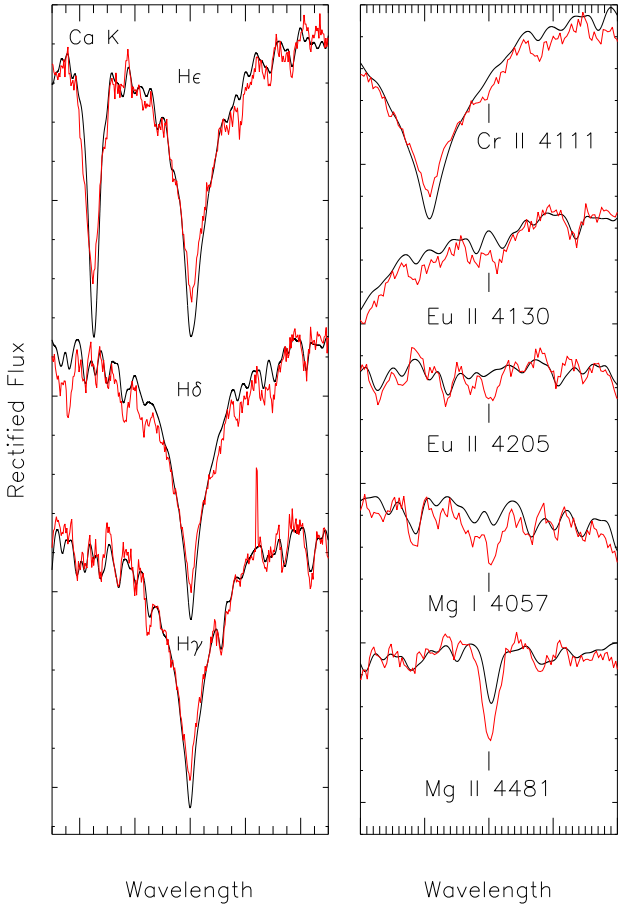}
\caption{Spectrum of J1940 (red) compared to a stellar model of $T_{\rm eff}=7700$\,K, $\log g=4.0$ and solar metallically (black). The first column shows the fits to the Balmer lines to determine the effective temperature, while the second column highlights the lines discussed in the text.}
\label{fig:spec_comp}
\end{figure}

Fig.\,\ref{fig:spec_comp} shows a comparison of the stellar spectrum with a model calculated with the above parameters to a) show the fit to the Balmer lines, and b) demonstrate the overabundances discussed in the text. The model spectrum does not fit well the cores of the hydrogen lines -- this is a known phenomenon in the Ap stars where the wings and the cores of the Balmer lines cannot be fit by a single temperature \citep{cowley01}.

The rotation phases at which the spectra were obtained have been calculated relative to the first light maximum in the WASP data set, such that:
\begin{equation}
  \phi(E)=(245\,3867.8805\pm0.0008)+ (\mbox{9{\fd}5344}\pm\mbox{0{\fd}0012})\times E,
  \label{equ:rot}
\end{equation}
where $E$ is the number of rotation cycles elapsed since the reference time. The rotation period is derived in section\,\ref{WASP}. The rotation phases are shown in the last column of Table\,\ref{tab:spec}.

    
\section{Photometric analysis}
\label{sec:phot}

As previously stated, the pulsation in J1940 was discovered by \citet{holdsworth14a} after conducting a survey of the SuperWASP archive in a search for pulsating A stars. Here we provide a more detailed discussion of the discovery data, with the addition of further ground-based data obtained with the 1.0-m and 1.9-m telescopes of the South African Astronomical Observatory (SAAO), the 0.9-m Small and Moderate Aperture Research Telescope System (SMARTS) telescope at the Cerro Tololo Inter-American Observatory (CTIO) and the 0.9-m telescope of the Southeastern Association for Research in Astronomy (SARA) at CTIO as part of independent and WET observations.

\subsection{Discovery data}
\label{WASP}

SuperWASP is one of the leading ground-based projects in the search for transiting exoplanets. This project is a two-site, wide-field survey, with instruments located at the Observatorio del Roque de los Muchachos on La Palma (WASP-N) and the Sutherland Station of the SAAO \citep[WASP-S; ][]{pollacco06}. Each instrument consists of eight 200-mm f/1.8 Canon telephoto lenses backed by cooled $2048\times2048$\,pixel Andor CCDs which provide a field-of-view of $\sim$64\,deg$^2$, with a pixel size of 13.7\,arcsec. Observations are made through broadband filters covering a wavelength range of $4000-7000$\,\AA\, and consist of two consecutive $30$-s integrations at a given pointing, with pointings being revisited, typically, every $10$\,min. The data are reduced with a custom reduction pipeline \citep[see][]{pollacco06} resulting in a `WASP-$V$' magnitude which is comparable to the Tycho-$2$ $V_t$\, passband.  Aperture photometry is performed at stellar positions provided by the USNO-B$1.0$\, input catalogue \citep{monet03} for stars in the magnitude range $5<V<15$.

SuperWASP observed J1940 for five seasons, 2006, 2007, 2008, 2011 and 2012, with some multiple observations per season using different cameras. The data were extracted from the archive after the standard processing had been applied and then passed through a resistant mean algorithm to remove out-lying points to improve the quality of the periodogram (see \citet{holdsworth14a} for an example and details). After trimming the data, 31\,756 data points remained. The details of the WASP observations are shown in Table\,\ref{tab:wasp}; multiple observations per season are distinguished with a letter after the season.

\begin{table*}
    \caption{Details of the WASP observations, and the results of a non-linear least-squares fit of the pulsation frequency in each of the seasons. BJD is given as BJD-245\,0000.0.}
\label{tab:wasp}
  \begin{tabular}{lcrccc}
    \hline
    \multicolumn{1}{c}{Season} & BJD & \multicolumn{1}{c}{Length} & Number of & Frequency & Amplitude \\
                 & start & \multicolumn{1}{c}{(d)}        & points        &  (\cd)          & (mmag)    \\
    \hline
    2006a & 3860.4399 & 51.0142 & 1355 & $176.33926	\pm	0.00295	$&$	3.51	\pm	1.03 $ \\
    2006b & 3860.4866 & 153.8625 & 4909 & $176.39091	\pm	0.00035	$&$	4.94	\pm	0.47	$ \\
    2006c & 3916.3760 & 97.8694 & 2740 & $176.39159	\pm	0.00058	$&$	4.79	\pm	0.53	$ \\
    2007a & 4211.5015 & 167.8901 & 5231 & $176.38928	\pm	0.00030	$&$	4.97	\pm	0.46	$ \\
    2007b & 4221.5029 & 166.7656 & 4294 & $176.38889	\pm	0.00042	$&$	4.25	\pm	0.49	$ \\
    2008   & 4574.5054 & 41.0215 & 1449 & $176.67150	\pm	0.00344	$&$	4.86	\pm	1.31	$ \\
    2011a & 5664.4956 & 168.8818 & 4298 & $176.39131	\pm	0.00049	$&$	4.66	\pm	0.69	$ \\
    2011b & 5691.4707 & 154.8721 & 3723 & $176.38837	\pm	0.00055	$&$	4.08	\pm	0.61	$ \\
    2012a & 6033.4854 & 73.9150 & 2096 & $176.39310	\pm	0.00175	$&$	4.38	\pm	0.93	$ \\
    2012b & 6042.5073 & 64.8941 & 1661 & $176.39036	\pm	0.00160	$&$	5.41	\pm	0.87	$ \\
    All & 	 3860.4399 & 2246.9614 & 31\,756 &$176.38997\pm0.00001	$&$	3.67	\pm	0.21	$ \\
    \hline
  \end{tabular}
\end{table*}

\begin{figure}
\includegraphics[width=\linewidth]{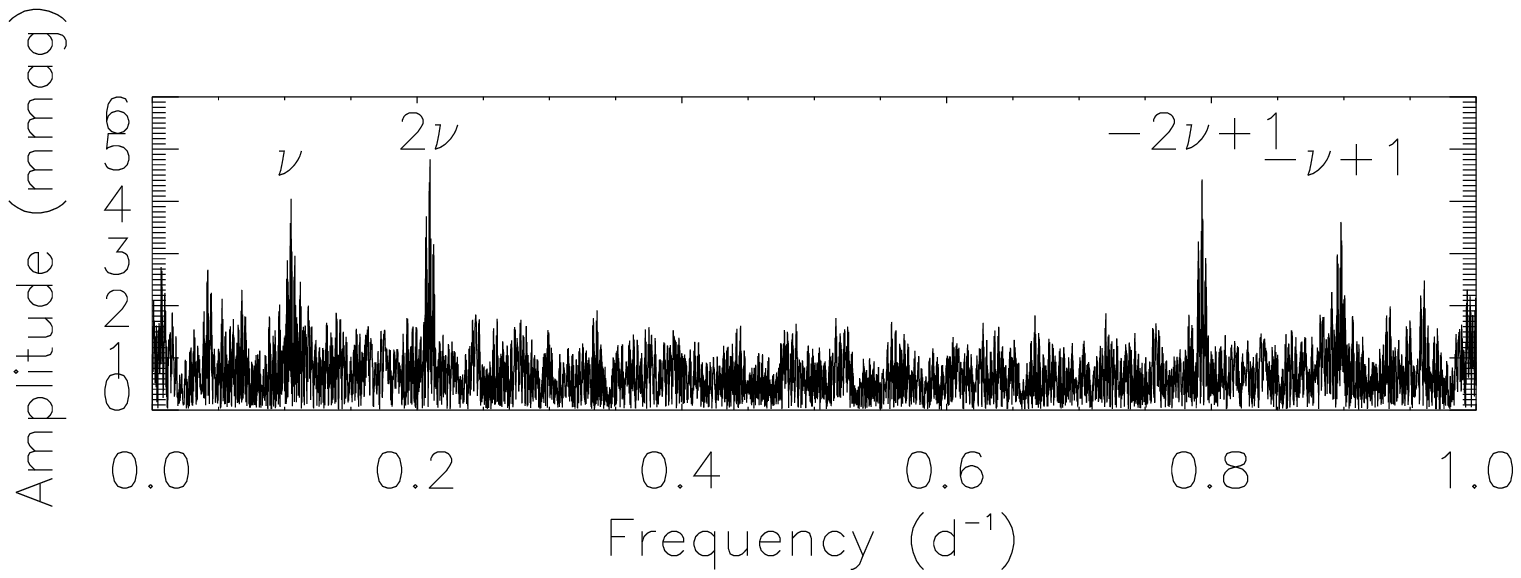}
\includegraphics[width=\linewidth]{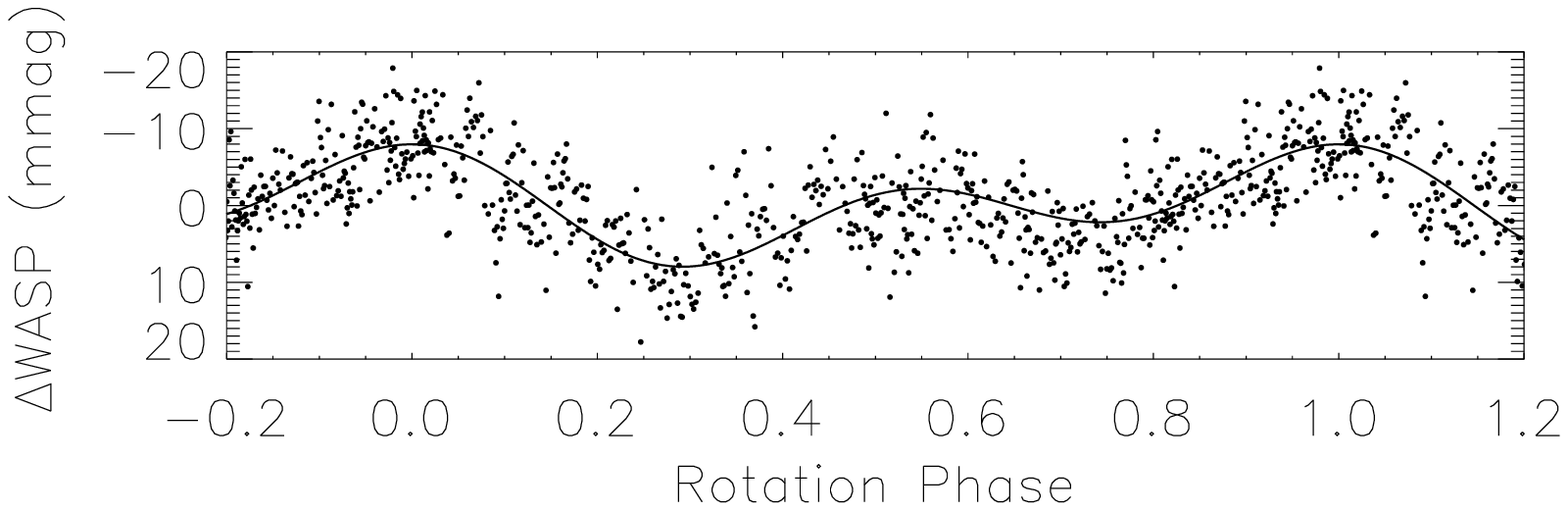}
  \caption{Top: low-frequency periodogram of the SuperWASP light curve showing the rotation frequency and its harmonic, as well as their aliases. Bottom: SuperWASP phase folded light curve on the frequency determined in the upper panel, i.e. $\nu=0.1049$\,\cd. The plot shows a clear double sinusoidal signature with a period of $1/\nu=9.5344$\,d. The data have been binned 50:1. }
  \label{fig:lc}
\end{figure}

A periodogram of the light curve shows a clear signature at low-frequency that is indicative of rotation (see Fig.\,\ref{fig:lc}). The stable spots on Ap stars are usually aligned with the magnetic axis, which in turn is inclined to the rotation axis, leading to the rigid rotator model of \citet{stibbs50}. Such a configuration results in brightness variations as the star rotates, thus allowing for the rotation period of the star to be determined. Using the {\sc{period}}04 program \citep{lenz05}, we detect a peak in the periodogram at a frequency of $0.20981\pm0.00001$\,\cd\, ($P=4.7662\pm0.0002$\,d), where the error is the analytical error given in {\sc{period}}04, using the method of \citet{montgomery99}. However, there is also a sub-harmonic to that frequency. We therefore simultaneously calculate the peak at $0.20981$\,\cd\ and a frequency at half that value, namely $0.10491$\,\cd. In doing so, we are able to determine a more precise measure of the rotation frequency and its harmonic. We find, $\nu_{\rm rot}=0.104884\pm0.000013$\cd\ and $2\nu_{\rm rot}=0.209809\pm0.000019$, corresponding to periods of $9.5344\pm0.0012$\,d and $4.7662\pm0.0004$\,d, respectively. By phase folding the data on the two frequencies independently, we are able to confirm that the longer rotation period is the correct period for the star. The phase folded light curve is shown in the bottom panel of Fig.\,\ref{fig:lc}.

\begin{figure}
  \includegraphics[width=\linewidth]{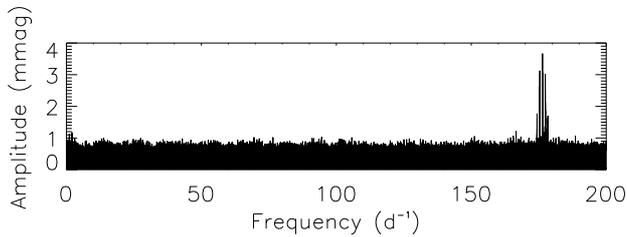}
  \caption{Periodogram of all seasons of WASP data. The data have been pre-whitened to 10\,\cd\, to the approximate noise level of the high-frequency range to remove the rotation signature and low-frequency noise. Note that the structure surrounding the pulsation signature is a result of daily aliases.}
  \label{fig:ft}
\end{figure}

The pulsation signature of J1940 is apparent in all seasons of data, and is shown in Fig.\,\ref{fig:ft}. Each season of data has been pre-whitened, to 10\,\cd, to the approximate noise level of the high-frequency range to remove the rotation signature and the remaining low-frequency `red' noise after the data have been processed by the WASP pipeline \citep{smith06}. The pulsation is clearly seen at a frequency of 176.39\,\cd\, with an amplitude of 3.67\,mmag in the WASP passband. The frequencies, amplitudes and phases of a non-linear least-squares fit for each season are provided in Table\,\ref{tab:wasp}.

Due to the survey nature of this ground-based data, not much further information can be extracted: the noise level in the high-frequency range is too great to discern, with confidence, sidelobes of the pulsation split by the rotation frequency. 

It must be noted here that the amplitudes are those as seen through the broadband filter of the WASP instrument (i.e. $4000-7000$\,\AA). Typically, roAp stars are observed with $B$-band filters where the signal-to-noise ratio is greatest for the observed pulsations. The amplitude suppression in the WASP filter is expected to be of the order 3 (following \citealt{medupe98}), demonstrating that J1940 is the highest amplitude roAp star known to date, and that it is a prime target for detailed follow-up observations using a $B$-band filter.


\subsection{Follow-up observations}
\subsubsection{SAAO data}

During an observing run at the SAAO in 2014 October/November, J1940 was observed as one of three roAp stars discovered in the WASP archive. Observations were made with both the 1.0-m and 1.9-m telescopes with the Sutherland High Speed Optical Cameras \citep[SHOC; ][]{coppejans13}, through a $B$-band filter. Integration times were 10\,s with a very short readout time (6.7\,ms) from frame-to-frame resulting in a cadence of 10.0067\,s. A log of the observations is presented in Table\,\ref{tab:log}.  

\begin{table*}
    \caption{Details of the SAAO, CTIO and SARA observations of J1940. BJD is given as BJD-245\,0000.0.}
\label{tab:log}
  \begin{tabular}{llcrccrr}
    \hline
     Year & \multicolumn{1}{c}{Date} & BJD & \multicolumn{1}{c}{Length} & Number of & Integration & \multicolumn{1}{c}{Telescope} & \multicolumn{1}{c}{Observer} \\ 
             &    			  & start & \multicolumn{1}{c}{(min)}        & points        &  time (s)	  &		     &		        \\
    \hline
    2014	\\
    		& Oct 29/30	& 6960.2529	& 125.7	& 752	& 	10	&	SAAO 1.9-m	&	DLH/BL \\
		& Oct 30/01	& 6961.2474	& 118.5	& 694	&	10	&	SAAO 1.9-m	&	DLH/BL \\		
		& Nov 05/06	& 6967.2422	& 167.2	& 1000	&	10	&	SAAO 1.9-m	&	DLH/BL \\		
		& Nov 06/07	& 6968.2432	& 166.6	& 999	&	10	&	SAAO 1.9-m	&	DLH/BL \\		
		& Nov 07/08	& 6969.2428	& 163.2	& 969	&	10	&	SAAO 1.9-m	&	DLH \\		
		& Nov 09/10	& 6971.2500	& 160.7	& 933	&	10	&	SAAO 1.9-m	&	DLH \\		
		& Nov 13/14	& 6975.2534	& 95.0	& 570	&	10	&	SAAO 1.0-m	&	DLH \\		
		& Nov 15/16	& 6977.2578	& 98.9	& 575	&	10	&	SAAO 1.0-m	&	DLH \\		
		& Nov 16/17	& 6978.2676	& 44.6	& 269	&	10	&	SAAO 1.0-m	&	DLH \\		
		& Nov 17/18	& 6979.2540	& 101.4	& 588	&	10	&	SAAO 1.0-m	&	DLH \\		
		& Nov 18/19	& 6980.2706	& 73.2	& 439	&	10	&	SAAO 1.0-m	&	DLH \\
   2015	\\
   		& Jul 15/16	& 7219.4399	& 307.4	& 1696	&	10	&	SAAO 1.0-m 	&	BL \\
   		& Jul 22/23	& 7226.3249	& 37.4	& 210	&	10	&	SAAO 1.9-m	&	BL \\
   		& Jul 26/27	& 7230.4912	& 261.6	& 1464	&	10	&	SAAO 1.9-m	&	BL \\
   		& Jul 27/28	& 7231.4458	& 273.2	& 1607	&	10	&	SAAO 1.9-m	&	BL \\
   		& Jul 28/29	& 7232.2808	& 302.0	& 1791	&	10	&	SAAO 1.9-m	&	BL \\
   2016	\\
   		& Jun 01/02	& 7541.4087	& 411.3	& 1217	&	20	&	SAAO 1.0-m	&	DLH/RRS \\
		& Jun 02/03	& 7542.4336	& 377.3	& 1132	&	20	&	SAAO 1.0-m	&	DLH/RRS \\
		& Jun 05/06	& 7545.3906	& 451.7	& 2626	&	10/20&	SAAO 1.0-m	&	DLH/RRS \\
		& Jun 06/07	& 7546.3940	& 446.0	& 2205	&	12	&	SAAO 1.0-m	&	DLH/RRS \\
		& Jun 07/08	& 7547.3867	& 446.1	& 2651	&	10	&	SAAO 1.0-m	&	DLH/RRS \\
		& Jun 08/09	& 7547.6904	& 378.4	& 309	&	30	&	CTIO 0.9-m	&	JLP \\
		& Jun 08/09	& 7548.4053	& 353.8	& 1982	&	10	&	SAAO 1.9-m	&	DLH \\
		& Jun 09/10	& 7548.6665	& 249.8 	& 232	&	20	&	CTIO 0.9-m	&	JLP \\
		& Jun 09/10	& 7549.3975	& 128.4	& 528	&	10	&	SAAO 1.9-m	&	DLH \\
		& Jun 10/11	& 7549.6938	& 354.8	& 271	&	30	&	CTIO 0.9-m	&	JLP \\
		& Jun 11/12	& 7550.6431	& 431.7	& 236	&	30	&	CTIO 0.9-m	&	JLP \\
		& Jun 11/12	& 7551.4204	& 158.3	& 635	&	10	&	SAAO 1.9-m	&	DLH \\
		& Jun 12/13	& 7551.6440	& 238.3	& 194	&	30	&	CTIO 0.9-m	&	JLP \\
		& Jun 12/13	& 7552.3892	& 214.6	& 985	&	10	&	SAAO 1.9-m	&	DLH \\
		& Jun 14/15	& 7554.5815	& 276.0	& 93		&	60	&	SARA 0.9-m	&	VP/HT/CLF \\
		& Jun 15/16	& 7555.5542	& 507.4	& 249	&	60	&	SARA 0.9-m	&	VP/HT/CLF \\
		& Jun 16/17	& 7556.3639	& 481.7	& 2874	&	10	&	SAAO 1.0-m	&	DLH \\
		& Jun 16/17	& 7556.6560	& 215.7	& 108	&	60	&	SARA 0.9-m	&	VP/HT/CLF \\
		& Jun 17/18	& 7557.3623	& 494.6	& 2953	&	10	&	SAAO 1.0-m	&	DLH \\
		& Jun 17/18	& 7557.8264	& 496.2	& 263	&	60	&	SARA 0.9-m	&	VP/HT/CLF \\
		& Jun 18/19	& 7558.3751	& 224.1	& 1334	&	10	&	SAAO 1.0-m	&	DLH \\
		& Jun 21/22	& 7561.3438	& 287.6	& 1723	&	10	&	SAAO 1.0-m	&	DLH \\
		& Jun 28/29	& 7568.3452	& 505.3	& 2998	&	10	&	SAAO 1.0-m	&	DLH \\
		 					
    \hline
  \end{tabular}
\end{table*}

The data were reduced following the pipeline outlined in \citet{provencal12}. Basic image reduction and aperture photometry was accomplished through the {\sc maestro} photometry pipeline described by \citet{dalessio10}.  Each image was corrected for bias and thermal noise, and normalised by its flat field. {\sc maestro} automatically covers a range of aperture sizes for the target and comparison stars. For each individual run we chose the combination of aperture size and comparison star(s) resulting in the highest quality light curve. 

The second step in data reduction was accomplished using the {\sc wqed}  pipeline \citep{thompson09}.  {\sc wqed} examines each light curve for photometric quality, removes outlying points, divides by suitable comparison stars, and corrects for differential extinction. Since we rely on relative photometry through the use of nearby comparison stars, our observational technique is not sensitive to oscillations with periods longer than a few hours. The final product from the {\sc wqed} pipeline is a series of light curves with times in seconds and amplitude variations represented as fractional intensity (ppt). We convert our light curves into units of magnitudes (where 1\,ppt\,=\,1.086\,mmag) and present our periodograms and analysis in units of mmag. Our final reduction step is to combine the individual light curves and apply barycentric corrections to create a complete light curve for the entire data set. 

\begin{figure}
  \includegraphics[width=\linewidth]{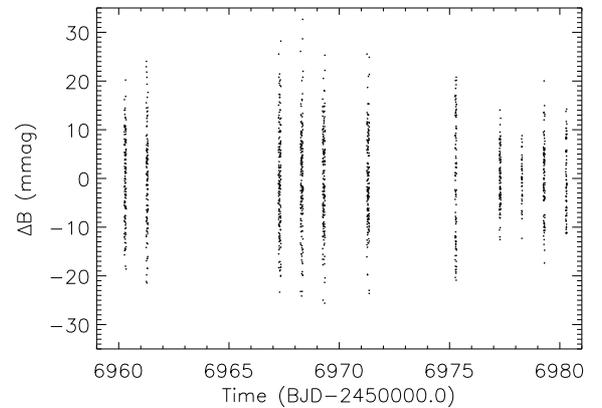}
  \caption{Light curve of the 2014 data from the SAAO, where the data have been binned to 60\,s integrations. Clearly evident is the night-to-night pulsational amplitude variation.}
  \label{fig:SAAO-2014}
\end{figure}

The light curves for each night of data are shown in Fig.\,\ref{fig:SAAO-2014}. Clearly evident is the night-to-night amplitude modulation of the variability as the star rotates. This will be discussed in more detail in section\,\ref{sec:wet}.

\begin{figure}
  \includegraphics[width=\linewidth]{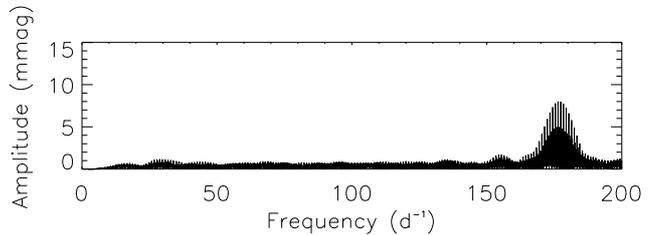}
  \caption{Periodogram of the 2014 SAAO data set. The pulsation is clearly present at 176.39\,\cd\, with an amplitude of 8.00\,mmag. The structure surrounding the peak are daily aliases.}
  \label{fig:2014-ft}
\end{figure}

We analyse the pulsation in the 2014 season of data using the 10\,s integration data set. Due to the method that was used to reduce the data, as described above, there are no low frequencies from the rotation or noise to be removed from the light curve. Fig.\,\ref{fig:2014-ft} shows a periodogram of the full 2014 data set. To extract the frequency, we perform a non-linear least-squares fit to the light curve, the results of which are shown in Table\,\ref{tab:2014-nlls}. The pulsation is extracted at a frequency of $176.3885\pm0.0015$\,\cd\, and an amplitude of $10.14\pm0.77$\,mmag. In addition to the pulsation, we are able to extract four sidelobes which fall at $\pm1$ and $\pm2\nu_{\rm rot}$. Such a signature suggests that J1940 is a quadrupole pulsator.

\begin{table*}
  \caption{The results of a non-linear least-squares fit to the 2014 light curve. The last column shows the difference in the frequency between that line and the previous. The zero-point for the phases is BJD-245\,6970.2871.}
  \label{tab:2014-nlls}
  \begin{tabular}{lcrrc}
    \hline
    ID & Frequency & \multicolumn{1}{c}{Amplitude} & \multicolumn{1}{c}{Phase} & Frequency difference\\
       & (\cd)     & \multicolumn{1}{c}{(mmag)}    & \multicolumn{1}{c}{(rad)} & (\cd) \\
    \hline
$\nu-2\nu_{\rm rot}$ 	&$	176.1699	\pm	0.0036	$&$	3.95	\pm	0.58	$&$	-1.513\pm	0.146$	\\				
$\nu-1\nu_{\rm rot}$ 	&$	176.2787	\pm	0.0044	$&$	4.68	\pm	0.91	$&$	1.838\pm0.149	$&$0.1088\pm	0.0057$ \\
$\nu$              		&$	176.3885	\pm	0.0015	$&$	10.14\pm	0.77	$&$	-0.846\pm0.055	$&$0.1099\pm	0.0047$ \\
$\nu+1\nu_{\rm rot}$	&$	176.5329	\pm	0.0374	$&$	1.01	\pm	0.34	$&$	-2.797\pm0.412	$&$0.1444\pm	0.0375$ \\
$\nu+2\nu_{\rm rot}$	&$	176.5992	\pm	0.0132	$&$	2.51	\pm	0.43	$&$	-2.621\pm0.159	$&$0.0663\pm	0.0397$ \\
\hline
  \end{tabular}
\end{table*}

\begin{figure}
  \includegraphics[width=\linewidth]{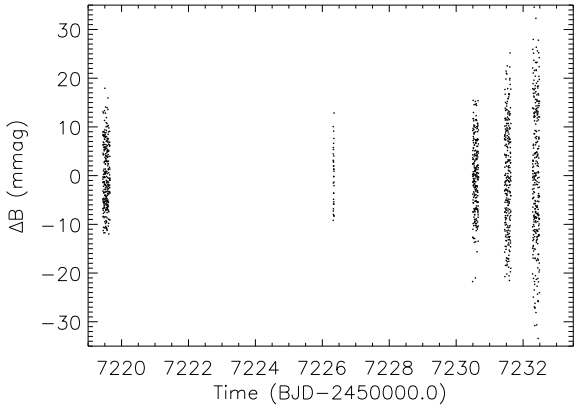}
  \caption{Light curve of the 2015 data from the SAAO, where the data have been binned to 60\,s integrations. Clearly evident is the night-to-night pulsational amplitude variation.}
  \label{fig:SAAO-2015}
\end{figure}

After the confirmation of the high-amplitude in J1940, it was clear that that star warranted another observing run. In 2015, J1940 was awarded two weeks of telescope time at SAAO and used the same telescopes/instrument and observing strategy as in 2014. However, poor weather led to just five nights of useful data. The observing log for 2015 can be seen in Table\,\ref{tab:log}. The nights which were clear yielded good data, as shown in Fig.\,\ref{fig:SAAO-2015}. 

\begin{figure}
  \includegraphics[width=\linewidth]{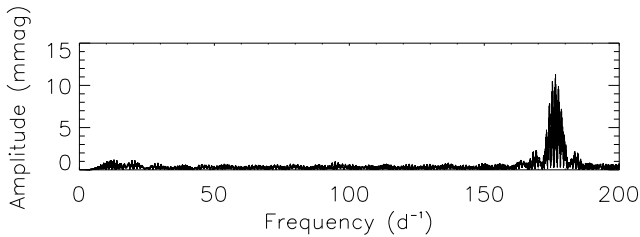}
  \caption{Periodogram of the 2015 SAAO data set. The pulsation is clearly present at 176.39\,\cd with an amplitude of 11.29\,mmag.}
  \label{fig:2015-ft}
\end{figure}

The periodogram of the 2015 data is shown in Fig.\,\ref{fig:2015-ft} (available online). A non-linear least-squares fit to the data is shown in Table\,\ref{tab:2015-nlls} (available online), identifying the pulsation at $176.3919\pm0.0007$\,\cd\, at an amplitude of $11.02\pm0.26$\,mmag. Due to the short run and low duty cycle of this data, we were unable to detect the rotational sidelobes.

\begin{table}
\centering
  \caption{The results of a non-linear least-squares fit to the 2015 light curve. The zero-point for the phase is BJD-245\,7225.9650.}
  \label{tab:2015-nlls}
  \begin{tabular}{lccr}
    \hline
    ID & Frequency & Amplitude & \multicolumn{1}{c}{Phase} \\
       & (\cd)     & (mmag)    & \multicolumn{1}{c}{(rad)} \\
    \hline
$\nu$	&	$176.3919\pm0.0007$	&	$11.02\pm0.26$	&	$0.375\pm0.026$ \\
\hline
  \end{tabular}
\end{table}

These data will be combined with the WET observations and analysed in section\,\ref{sec:all_data}.


\subsubsection{WET observations}
\label{sec:wet}

The final observations we present here are from a Whole Earth Telescope campaign in June 2016. The campaign secured three weeks of telescope time at SAAO, six nights at CTIO, and four nights at SARA. A log of the observations is shown in Table\,\ref{tab:log}.

The SAAO data were collected as described above, using both the 1.0-m and 1.9-m telescopes. The CTIO data were collected using the 0.9-m telescope backed by the Tek (STIe) $2048\times2048$ pixel CCD (\#3). The total readout time between frames is 42.9\,s. The SARA data were collected with the 0.9-m telescope, backed by a $2048\times2048$ ARC-E2V42-40 chip. The readout time between frames is 2.9\,s. The readout times quoted are in addition to the integration times shown in Table\,\ref{tab:log}, thus the typical cadence for CTIO observations is 72.9\,s and 62.9\,s for SARA observations\footnote{We note that the SARA observations are not continuous due to multiple targets observed during the nights.}.

All observations were made through a $B$-band filter to normalise wavelength response and minimise extinction effects. The data were reduced following the steps outlined above. We assume that our target oscillates around a mean light level. This important assumption allows us to assess overlapping light curves from different telescopes and identify and correct any residual vertical offsets that are instrumental in nature. The treatment of overlapping data is discussed in detail in \citet{provencal09}. We find no significant differences between the noise level in the periodograms using: 1) the combination of every light curve including overlapping segments from different telescopes, 2) the combination of light curves where we retain the high signal to noise observations in overlapping segments and 3) combining all light curves incorporating data weighted by telescope aperture.

\begin{figure}
  \includegraphics[width=\linewidth]{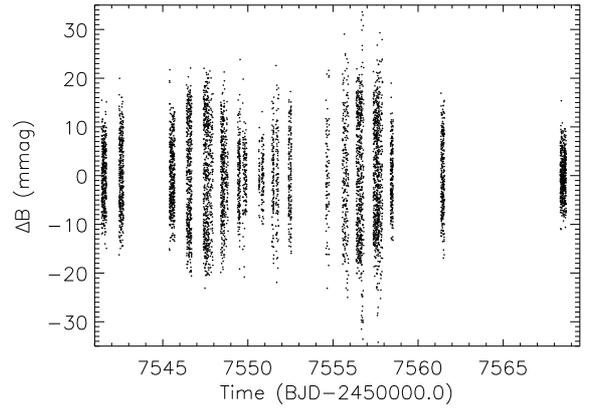}
  \caption{Light curve of the 2016 data from the WET, where the data have been binned to 60\,s integrations. Clearly evident is the night-to-night pulsational amplitude variation.}
  \label{fig:wet-lc}
\end{figure}

Due to the different instrument responses, telescope apertures, and observing conditions, different integration times were used throughout the campaign. As such, data used for the analysis have been binned to the equivalent of 60\,s integrations. The light curve of the observations is shown in Fig.\,\ref{fig:wet-lc}.

The duty cycle of the WET observations is not optimal at just 21\,per\,cent. For the consecutive nights where SAAO, CTIO and SARA time was awarded, the duty cycle becomes 23\,per\,cent -- during this period continuous observing from dusk at SAAO to dawn at SARA provided the least fragmented data set. These duty cycles are much lower compared to the most well studied roAp star with the WET, namely HR\,1217 \citep{kurtz05}, which achieved a duty cycle of 36\,per\,cent. However, even with a low duty cycle, multisite observations still serve to reduce daily aliases which strongly affect single site observations. The data set presented here is the best available for J1940 to date.

\begin{figure}
  \includegraphics[width=\linewidth]{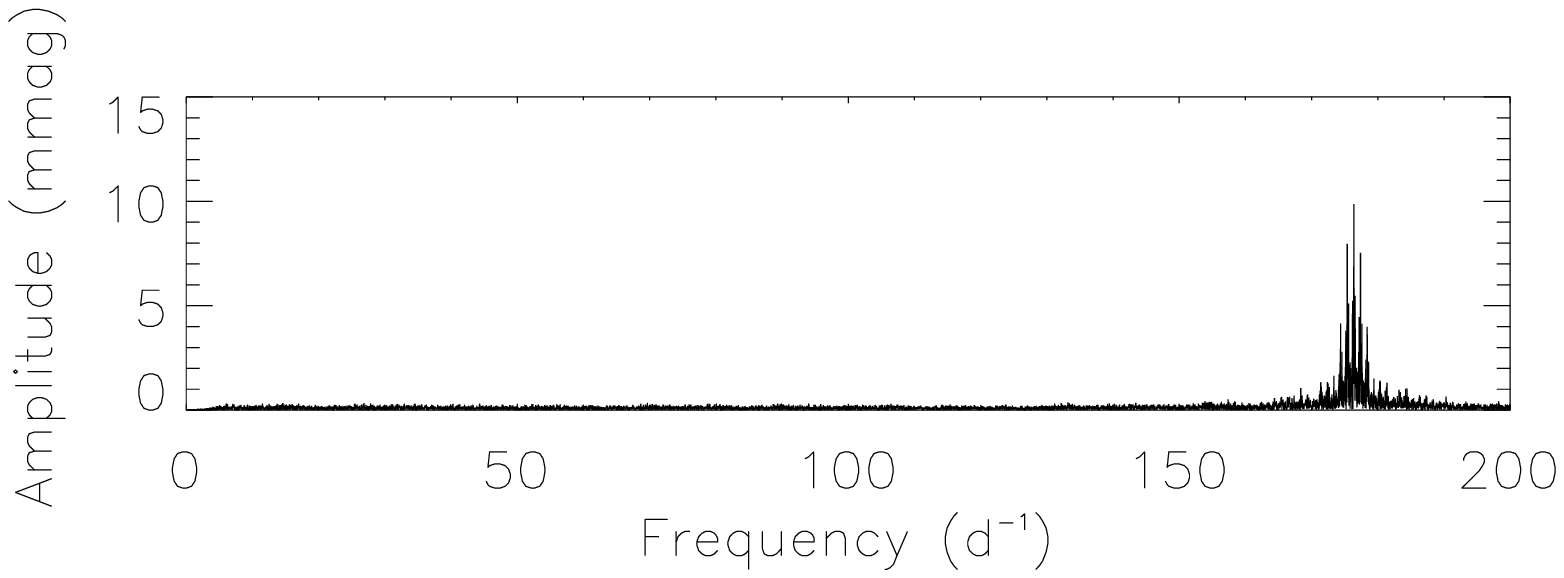}
  \includegraphics[width=\linewidth]{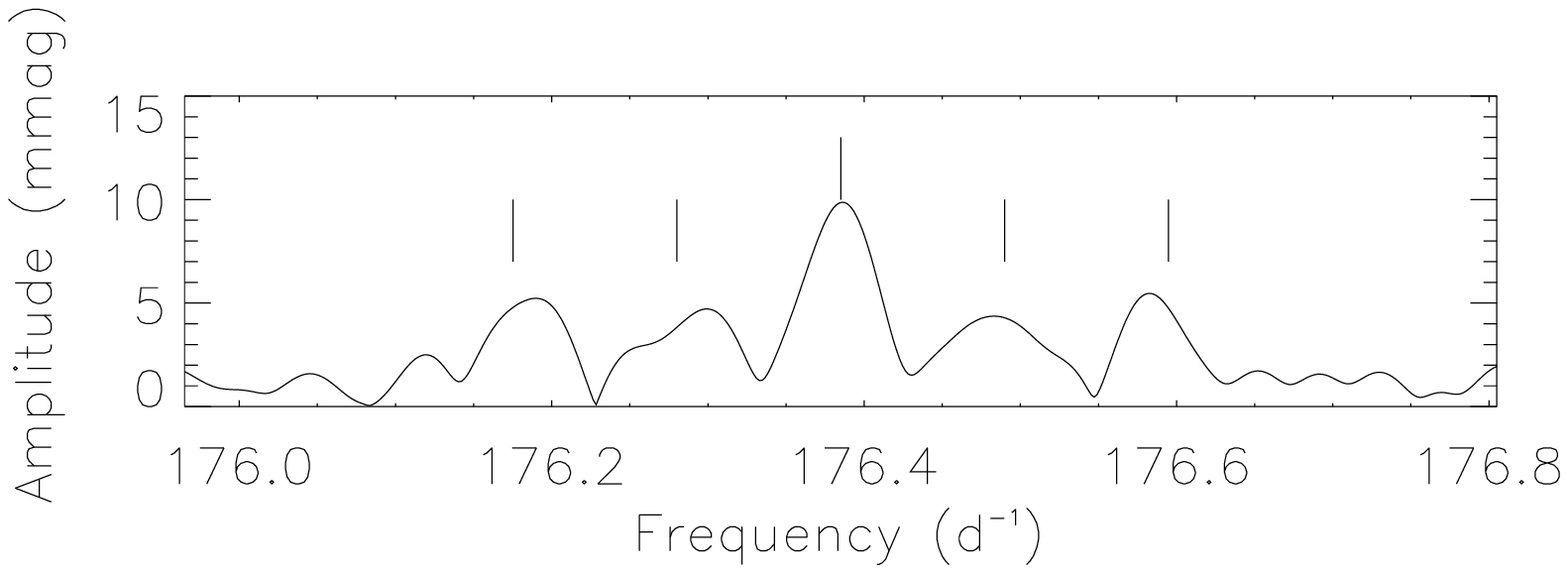}
  \caption{Top: periodogram of the full WET data set, binned to 60\,s. Bottom: detailed view of the pulsation, showing the main peak and four rotationally split sidelobes. The vertical bars indicate the positions of sidelobes split by exactly the rotation frequency.}
  \label{fig:2016-ft}
\end{figure}

We calculate a periodogram of the WET data which is shown in Fig.\,\ref{fig:2016-ft}. It is clear from the lower panel in the plot that the rotational sidelobes are well resolved. This is also evident in Table\,\ref{tab:wet-nlls}, which shows the result of a non-linear least-squares fit to the light curve. The presence of four rotationally split sidelobes confirms that J1940 is a quadrupole pulsator.

\begin{table*}
  \caption{The results of a non-linear least-squares fit to the WET light curve. The last column shows the difference in the frequency between that line and the previous. The zero-point for the phases is BJD-245\,7555.0521.}
  \label{tab:wet-nlls}
  \begin{tabular}{lccrc}
    \hline
    ID & Frequency & Amplitude & \multicolumn{1}{c}{Phase} & Frequency difference\\
       & (\cd)     & (mmag)    & \multicolumn{1}{c}{(rad)} & (\cd) \\
    \hline

$\nu-2\nu_{\rm rot}$ 	&$	176.1752	\pm	0.0012	$ & $	2.10	\pm	0.12	$ & $	-1.087	\pm	0.065	$ \\				
$\nu-1\nu_{\rm rot}$ 	&$	176.2787	\pm	0.0011	$ & $	2.30	\pm	0.13	$ & $	-2.297	\pm	0.063	$ & $	0.1035	\pm	0.0016	$ \\
$\nu$              		&$	176.3850	\pm	0.0003	$ & $	8.45	\pm	0.13	$ & $	-2.940	\pm	0.015	$ & $	0.1062	\pm	0.0011	$ \\
$\nu+1\nu_{\rm rot}$	&$	176.4916	\pm	0.0018	$ & $	1.36	\pm	0.12	$ & $	2.016	\pm	0.109	$ & $	0.1066	\pm	0.0019	$ \\
$\nu+2\nu_{\rm rot}$	&$	176.5931	\pm	0.0011	$ & $	2.58	\pm	0.12	$ & $	0.498	\pm	0.059	$ & $	0.1016	\pm	0.0021	$ \\

\hline
  \end{tabular}
\end{table*}

Although we are able to continue fitting and extracting higher and higher order sidelobes, the S/N of the peaks decreases as a result of the increasing cross-talk with the aliased sidelobes. Therefore, although we believe that the high order sidelobes are real and present, we stop our extraction at $\pm2\nu_{\rm rot}$ where the S/N$\geq4$ for each peak (as shown in Table\,\ref{tab:wet-nlls}) and the frequencies are close to what is calculated using the rotation period.

\begin{figure*}
  \includegraphics[width=\textwidth]{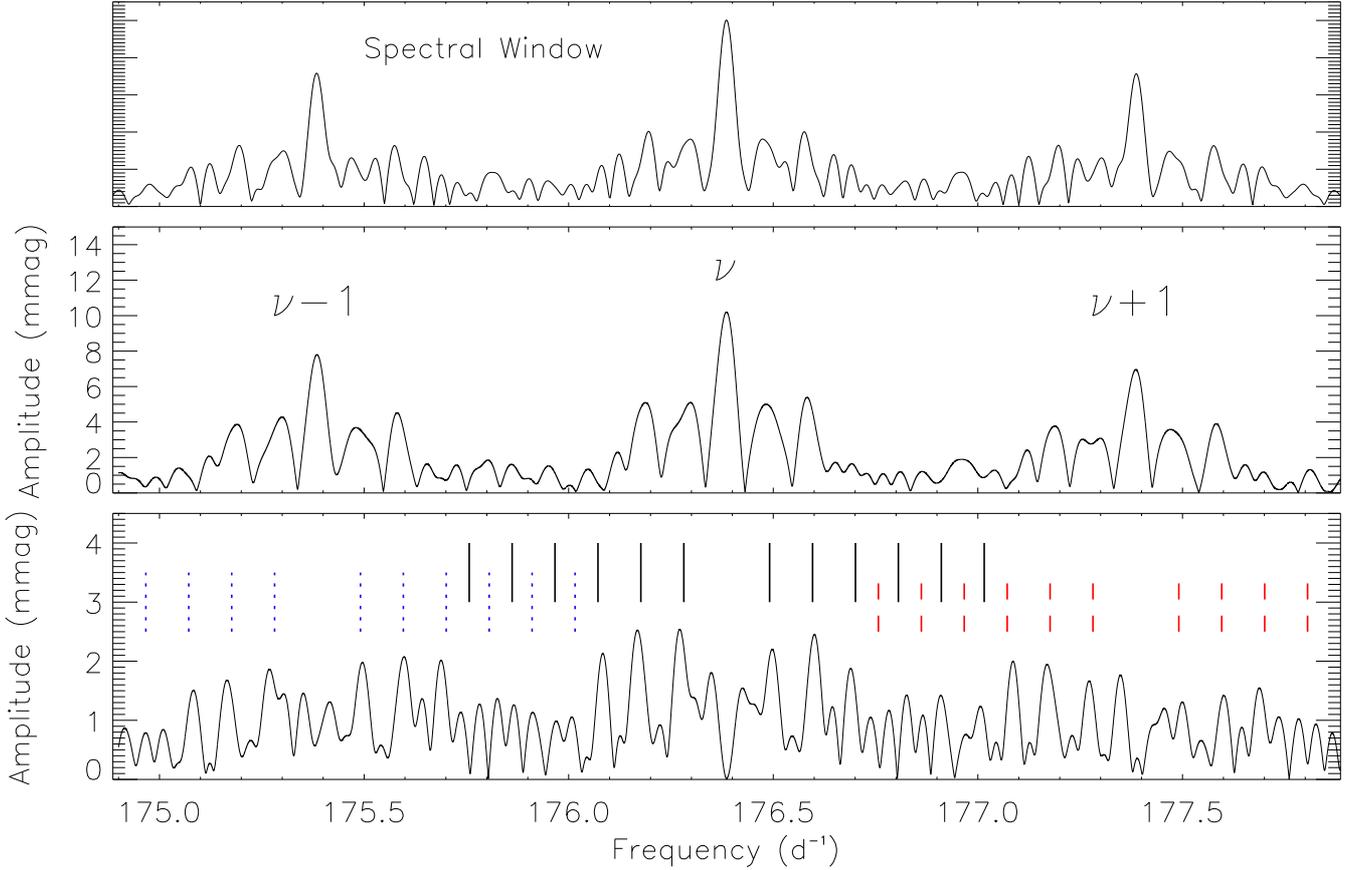}
  \caption{Top: spectral window function of the WET data, folded on the pulsation frequency. Middle: the periodogram of the WET data showing the pulsation at $\nu$ and the daily aliases at $\nu\pm1$. Bottom: periodogram of the WET data after the main pulsation frequency has been prewhitened. The solid vertical bars indicate the position of the rotation sidelobes of the pulsation frequency $\nu$, and the location of the sidelobes of the positive (red dashed bars) and negative (blue dotted bars) aliases. Where the sidelobes overlap, the errors on the `true' extracted sidelobes increase as a result of cross-talk between the two signals. Only a continuous dataset will alleviate this problem.} 
  \label{fig:sidelobes_alias}
\end{figure*}

As we increase the number of sidelobes that we fit, the error in the amplitude increases. This is a result of the window pattern and cross-talk between the aliases of the sidelobes and the sidelobes we are fitting. As we fit more and more sidelobes, we increase the distance, in frequency, from the central peak. As a result, the frequencies we are fitting start to overlap with the aliased sidelobes of $\pm\nu$. This is shown graphically in Fig.\,\ref{fig:sidelobes_alias}. The top panel shows the spectral window of the WET data, with the middle one showing the pulsation and its positive and negative aliases separated by $\pm1$\,\cd. In the bottom panel we indicate the position of the rotational sidelobes with solid bars and aliased sidelobes with coloured broken bars. As we fit frequencies $\pm3\nu_{\rm rot}$ and greater, we suffer from cross-talk with the aliased sidelobes, hence we do not include these sidelobes in our analysis.

The oblique pulsator model (OPM) expects that the sidelobes of the pulsation have the same amplitude (in $\pm n\nu_{\rm rot}$ pairs) and are exactly split by the rotation frequency of the star. Therefore, to test the OPM for J1940, we split the sidelobes by the derived rotation frequency and perform a linear least-squares fit to the data. When performing this test, we select the zero-point in time such that the phases of the first sidelobes are equal. The results of this are shown in Table\,\ref{tab:wet-lls}. For a pure quadrupole pulsator, we expect the phases of the quintuplet to be equal. Although the phases presented in Table\,\ref{tab:wet-lls} are almost equal, the average separation of the phases is greater than $6.5\,\sigma$. This result, combined with the presence of further rotational sidelobes and the unequal sidelobe amplitudes, leads us to conclude that J1940 is pulsating in a distorted mode. We will revisit this result in section\,\ref{sec:modelling}.

\begin{table}
\centering
  \caption{The results of a linear least-squares fit to the WET light curve. The zero-point for the phases chosen to be BJD-245\,7557.7219, to force the first sidelobes to have equal phase.}
  \label{tab:wet-lls}
  \begin{tabular}{lccrc}
    \hline
    ID & Frequency & Amplitude & \multicolumn{1}{c}{Phase}\\
       & (\cd)     & (mmag)    & \multicolumn{1}{c}{(rad)} \\
    \hline
$\nu-2\nu_{\rm rot}$ 	&$	176.1755	$ & $	2.15	\pm	0.11	$ & $	2.914	\pm	0.051	$ \\
$\nu-1\nu_{\rm rot}$ 	&$	176.2804	$ & $	2.26	\pm	0.12	$ & $	2.720	\pm	0.054	$ \\
$\nu$              		&$	176.3852	$ & $	8.47	\pm	0.12	$ & $	3.060	\pm	0.014	$ \\
$\nu+1\nu_{\rm rot}$	&$	176.4901	$ & $	1.37	\pm	0.12	$ & $	2.720	\pm	0.089	$ \\
$\nu+2\nu_{\rm rot}$	&$	176.5950	$ & $	2.58	\pm	0.11	$ & $	2.218	\pm	0.043	$ \\
\hline
  \end{tabular}
\end{table}

\subsubsection{All data}
\label{sec:all_data}

With the availability of three seasons of follow-up data, we are able to perform a much more precise analysis of the pulsations in J1940 as the frequency resolution is dependent on the time-base of the observations. To perform this analysis, we use data which is binned to 60\,s integrations so that each data point has the same weighting. 

As before, we perform linear and non-linear least-squares fitting of the data. The results of the non-linear fits are shown in Table\,\ref{tab:all-nlls}. To test the geometry of the star, we again force the phases of the first sidelobes to be equal, set the frequencies to be split exactly by the rotation frequency and perform a linear least-squares fit. The result of this is shown in Table\,\ref{tab:all-lls}. As with the WET data, we see that the phases of the quintuplet are not all equal, with the average separation being $11.4\,\sigma$. Using this data set with a longer time-base, we are able to confirm that the pulsation is J1940 is indeed distorted. 

\begin{table*}
  \caption{The results of a non-linear least-squares fit to all data. The last column shows the difference in the frequency between that line and the previous. The zero-point for the phases is BJD-245\,7270.3489.}
  \label{tab:all-nlls}
  \begin{tabular}{lccrc}
    \hline
    ID & Frequency & Amplitude & \multicolumn{1}{c}{Phase} & Frequency difference\\
       & (\cd)     & (mmag)    & \multicolumn{1}{c}{(rad)} & (\cd) \\
    \hline
$\nu-2\nu_{\rm rot}$ 	&$	176.173294	\pm	0.000025	$ & $	2.58	\pm	0.09	$ & $	-1.686	\pm	0.042	$ \\				
$\nu-1\nu_{\rm rot}$ 	&$	176.278217	\pm	0.000043	$ & $	1.55	\pm	0.10	$ & $	-2.187	\pm	0.072	$ & $	0.104923	\pm	0.000050	$ \\
$\nu$              		&$	176.383123	\pm	0.000008	$ & $	8.56	\pm	0.10	$ & $	-2.064	\pm	0.013	$ & $	0.104906	\pm	0.000044	$ \\
$\nu+1\nu_{\rm rot}$	&$	176.488089	\pm	0.000030	$ & $	2.16	\pm	0.10	$ & $	-2.314	\pm	0.052	$ & $	0.104966	\pm	0.000031	$ \\
$\nu+2\nu_{\rm rot}$	&$	176.592901	\pm	0.000033	$ & $	1.98	\pm	0.09	$ & $	-3.096	\pm	0.054	$ & $	0.104812	\pm	0.000045	$ \\
\hline
  \end{tabular}
\end{table*}

\begin{table}
\centering
  \caption{The results of a linear least-squares fit to all the data. The zero-point for the phases chosen to be BJD-245\,7270.3489.}
  \label{tab:all-lls}
  \begin{tabular}{lccrc}
    \hline
    ID & Frequency & Amplitude & \multicolumn{1}{c}{Phase}\\
       & (\cd)     & (mmag)    & \multicolumn{1}{c}{(rad)} \\
    \hline
$\nu-2\nu_{\rm rot}$ 	&$	176.1734   $ & $	      2.55\pm    0.09 $ & $       -1.776  \pm  0.035$ \\
$\nu-1\nu_{\rm rot}$ 	&$	176.2783   $ & $	      1.62\pm    0.10 $ & $       -2.276  \pm  0.057$ \\
$\nu$              		&$	176.3831   $ & $	      8.54\pm    0.10 $ & $       -2.080  \pm  0.011$ \\
$\nu+1\nu_{\rm rot}$	&$	176.4880   $ & $	      2.10\pm    0.10 $ & $       -2.276  \pm  0.044$ \\
$\nu+2\nu_{\rm rot}$	&$	176.5929   $ & $	      1.99\pm    0.09 $ & $       -3.110  \pm  0.045$ \\
\hline
  \end{tabular}
\end{table}

Although the sidelobes at $\pm3\nu_{\rm rot}$ are present in the data, they are below the S/N$=4$ limit, so we do not include them in the analysis. Therefore, in fitting the entire data set, we only extract the central quintuplet.

Finally, we are able to identify the first harmonic of the pulsation at $2\nu$. At higher frequencies, the signal is lost in the noise. The presence of the harmonics demonstrate the non-sinusoidal nature of the pulsations seen in roAp stars. However, it is not yet clear what these harmonics can tell us about the pulsations in roAp stars.

\subsection{Testing amplitude and phase variability}

Some roAp stars show very stable pulsations while others show dramatic variability over the time span that they have been observed. \citet{kurtz94,kurtz97} discussed the frequency variability observed in HR\,3831 through the analysis of 16\,yr of ground based data. Work by \citet{martinez94} highlighted eight other roAp stars for which frequency variability has been detected. Recent studies at high photometric precision have shown significant phase/frequency variations in the roAp stars observed by {\it Kepler} \citep[e.g.][]{holdsworth14b,smalley15,holdsworth16}. 

\begin{figure}
  \includegraphics[width=\linewidth]{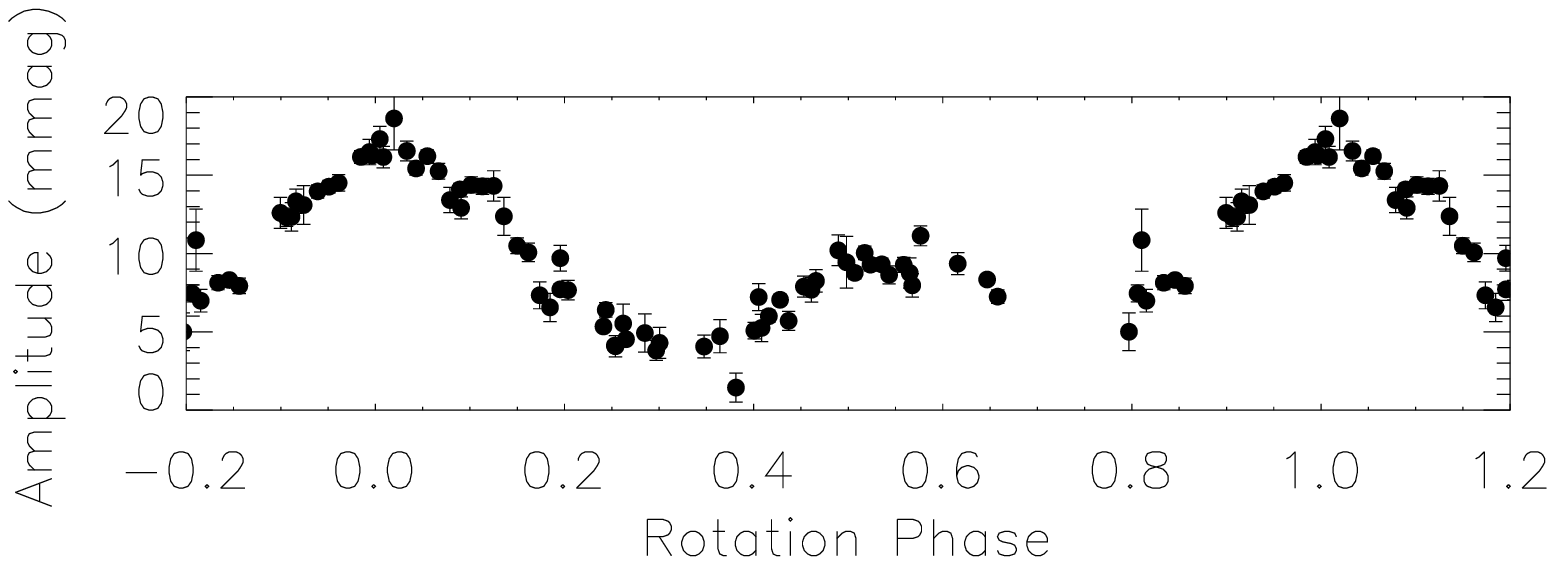}
  \includegraphics[width=\linewidth]{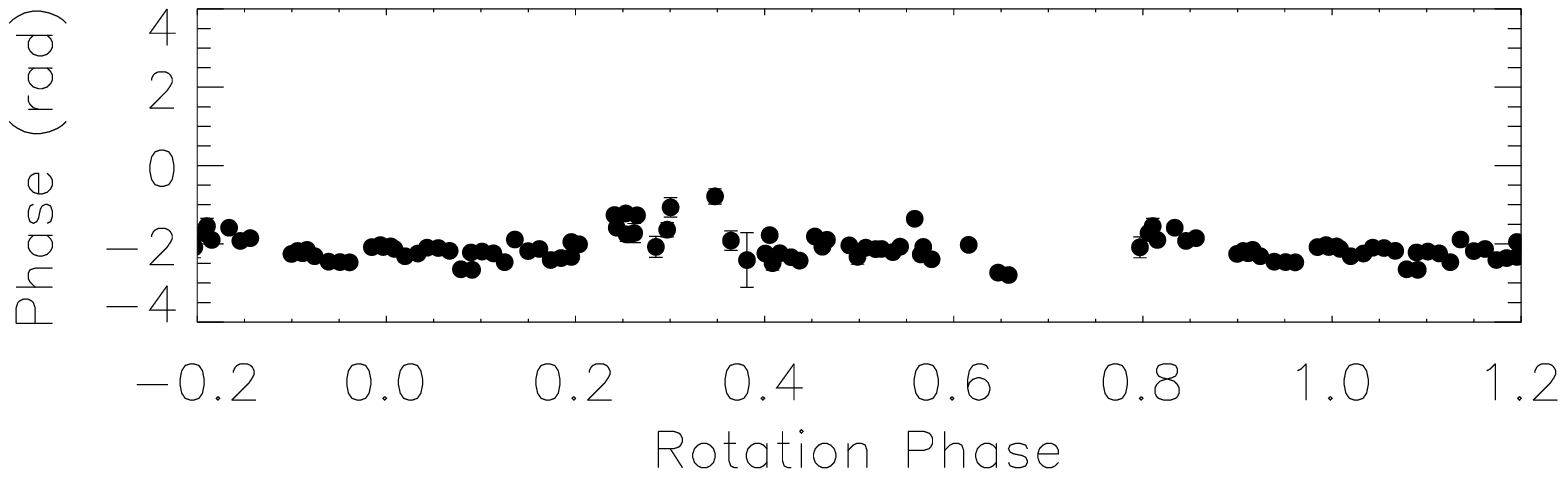}
  \caption{Top: pulsation amplitude variation with rotation phase. Bottom: pulsation phase variation with rotation phase. The zero-point for the rotation phase is chosen when the pulsation and most of the sidelobes are in phase.}
  \label{fig:all-ph-amp-phi}
\end{figure}

With the combined data set, which provides us nearly full rotational phase coverage of the star, we test the stability of the pulsation detected in J1940. To conduct this test, we split the data into sections of 20 pulsation cycles, or about 0.11\,d, and calculate the amplitude and phase at fixed frequency. As phase and frequency are inextricably intertwined, a slope in phase means a different frequency would provide a better fit. The results of this procedure are shown in Fig.\,\ref{fig:all-ph-amp-phi}.

Concentrating on the amplitude variations, for a non-distorted quadrupole mode, the OPM predicts three local maxima and minima in the pulsation amplitude as the star rotates, as a consequence of two nodes passing the line-of-sight (for the geometry of J1940; see section\,\ref{sec:modelling}). As can be seen, the amplitude of the pulsation does not behave as expected. There are only two minima in the phased plot, suggesting a distorted mode. Further confirmation of this is given as the amplitude never goes to zero, as would be expected as a node passes the line-of-sight for a pure $\ell=2$ mode. 

In Fig.\,\ref{fig:all-ph-amp-phi} we choose the zero-point for the rotation phase such that $\phi_{\rm rot}=0$ is the time at pulsational amplitude maximum. This is different from the phase at which light maximum occurs, which is a result of the surface spots on the star. We assume that the spots are at the magnetic poles, however this is not always the case \citep[e.g. ][]{luftinger10,kochukhov15}. The difference between the two epochs of maxima equates to $1.312\pm0.428$\,d (or $0.138\pm0.045$\,rotation periods). We can interpret this offset in three ways:

\begin{enumerate}

\item a lag maybe introduced if the rotation frequency is not precisely determined, as a result of the large gap between the zero-point in light maximum and the zero-point in pulsation maximum (which equates to $\sim$357 rotation cycles).  However, as the rotation period is derived over 235.7 cycles, we are confident that this has sufficient precision to provide an accurate time of light maximum during the follow-up observations. 

\item there is a significant longitudinal offset between the pulsation axis and the magnetic axis, which is where we assume the spots to form. In this case, the pulsation pole rotates into view before the spots. This scenario is not testable with current instrumentation due to the faintness of J1940. 

\item the pulsation and magnetic axes are closely aligned, but the spots are {\it not} concentric about the magnetic poles, as is demonstrated by some roAp stars \citep[e.g. HR\,3831;][]{kurtz00}. If this is the case, it does not affect the pulsation analysis we present here as we analyse integrated photometric observations over the entire photosphere, rather than radial velocity variations derived from specific elemental species. An alternative model to the oblique pulsator model was formulated by \citet{mathys85} where the pulsational light amplitude variations are created by the inhomogeneous distribution of the flux-to-radius variations caused by spotty abundance and temperature distributions in the photosphere. However, as can be seen in the theory of Mathys, the effect of the flux variations alone caused by the spots can only account for a change of about 2 per cent in the pulsation amplitude of J1940, whereas we actually measure a change on the order of 500 per cent, thus making the spot position a negligible effect.

\end{enumerate}

Unfortunately, we do not have the information here to firmly rule out any of these scenarios. Observations of J1940 by the {\it TESS} mission will be able to shed light on the first scenario, with the second and third only testable with substantially larger telescopes than are currently available, or unfeasibly long exposure times ($\sim\,10\,000$\,s) with current instruments.

\begin{figure*}
  \includegraphics[width=\textwidth]{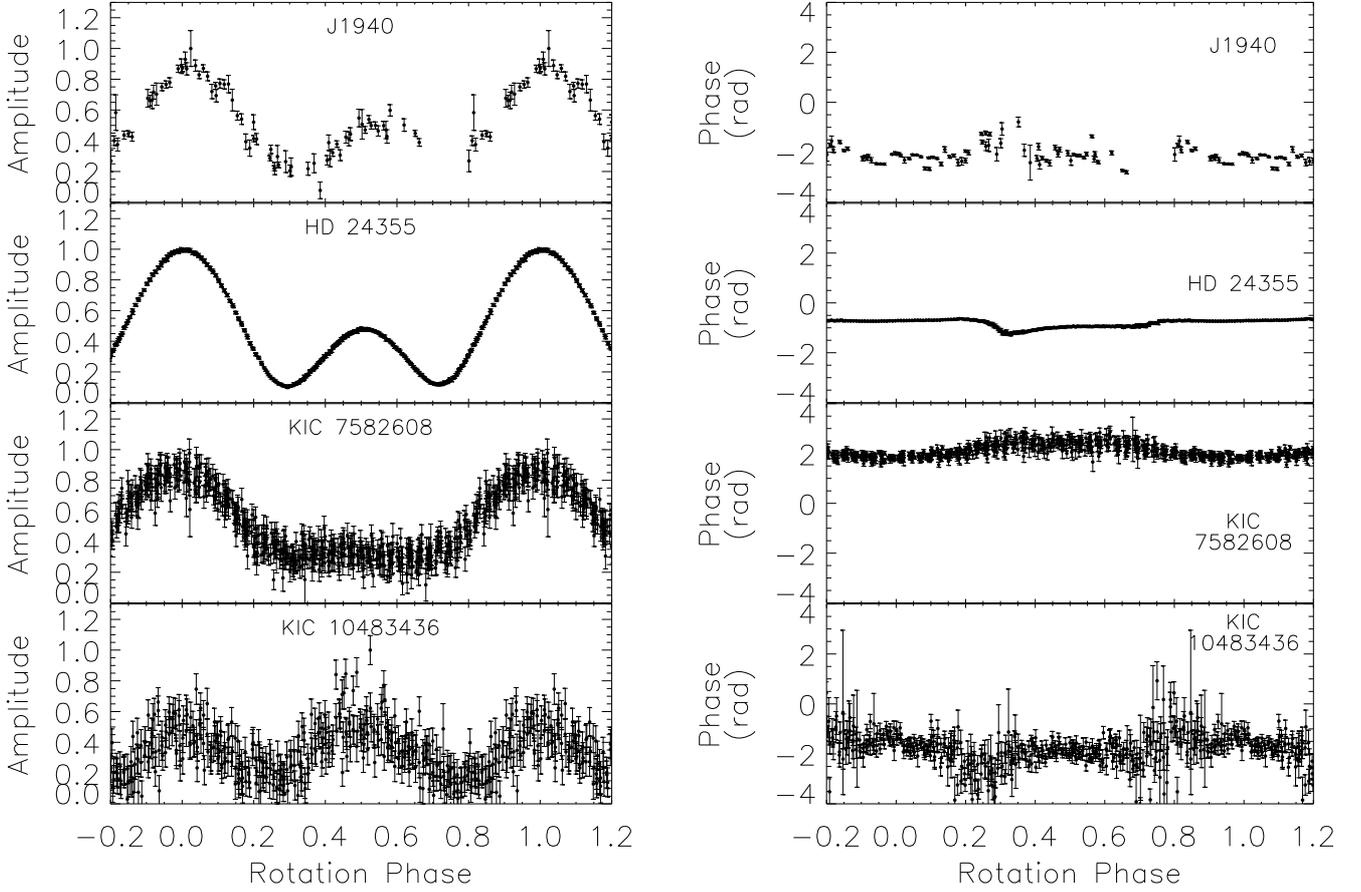}
  \caption{The amplitude and phase variations of other quadrupole roAp stars studied in detail. All stars show an amplitude modulation reminiscent of dipole modes rather than quadrupole modes. The phase variations in these stars do not show the expected $\pi$-rad phase change at quadrature as is expected from the OPM. The amplitudes have been normalised to the maximum amplitude for each star.}
  \label{fig:other_quad}
\end{figure*}

We now consider the phase variations seen in Fig.\,\ref{fig:all-ph-amp-phi}. By folding the pulsation phase on the rotation period, we are able to see that the fitted frequency is the correct frequency for the length of the observations. If this were not the case, there would be a linear trend to the points. The single line, however, is not expected. According to the OPM, the phase should flip by $\pi$-rad when a node crosses the line-of-sight -- something which does not occur here. In fact, the pulsation phase is almost constant as the star rotates. This result is very similar to those seen in HD\,24355 \citep{holdsworth16}, KIC\,7582608 \citep{holdsworth14b} and KIC\,10483436 \citep{balona11b}, all quadrupole pulsators with distorted modes observed with the {\it Kepler} space telescope. We provide a comparison of the four pulsators in Fig.\,\ref{fig:other_quad}. Given the irregularities in the phase variations, we attempt to model the star to further understand this distorted pulsator in section\,\ref{sec:modelling}.

\subsection{Constraining the mode geometry}

We strive to derive the geometry of the star by comparing the amplitudes of the sidelobes, following the method of \citet{kurtz90}. Assuming that the quintuplet is representative of a quadrupolar pulsator, and assuming that we have an axisymetric non-distorted mode (which we have shown we do not), such that $\ell=2$ and $m=0$, then in the absence of limb darkening the following relation is applicable:

\begin{equation}
\tan i\tan\beta=4\frac{A^{(2)}_{-2}+A^{(2)}_{+2}}{A^{(2)}_{-1}+A^{(2)}_{+1}},
\label{eq:OPM}
\end{equation}
where $i$ is the inclination angle of the star, $\beta$ is the angle of obliquity between the rotation and the pulsation axes, and $A^{(2)}_{\pm1}$ and $A^{(2)}_{\pm2}$ are the amplitudes of the first and second sidelobe pairs of the quadrupole pulsator, respectively. Substituting values from Table\,\ref{tab:all-lls} into equation\,\ref{eq:OPM}, we derive that $\tan i\tan\beta=4.88\pm0.23$ for this simplified case. Although the assumptions made will affect our result, values derived from this relationship will provide the first steps in determining the geometry of this pulsator.

From the rotational light variations, we know that $i$ and $\beta$ sum to greater than $90^\circ$ as we see both magnetic poles (assuming that the spots causing the variations are at the magnetic poles, which may not be the case). Further to this, values of $i$ and $\beta$ near to 90$^\circ$ are not permitted for a non-distorted quadrupole mode, within the constraint of equation\,(\ref{eq:OPM}), because the line-of-sight would not pass over a node; the nodes of a pure quadrupole are at $\pm54.7^\circ$ (colatitude), since the Legendre polynomial is given as $P^0_2(\cos\theta)=(3\cos^2\theta -1)/2$, therefore $i-\beta$ must be less than $54.7^\circ$. Table\,\ref{tab:iandbeta} shows a set of values of $i$ and $\beta$ that satisfy all the constraints above.

\begin{table}
\centering
  \caption{The values of $i$ and $\beta$ which satisfy equation\,(\ref{eq:OPM}). We list values up to $i=\beta$, where the values of $i$ and $\beta$ reverse. The last two columns show the extreme values of the angle between the line-of-sight and the pulsation pole; these can be useful in visualising the pulsation geometry.}
  \label{tab:iandbeta}
  \begin{tabular}{rrrr}
    \hline
    \multicolumn{1}{c}{$i$} & \multicolumn{1}{c}{$\beta$} & \multicolumn{1}{c}{$i+\beta$} & \multicolumn{1}{c}{$i-\beta$} \\
    \hline
83.6 & 28.9 & 112.5 & 54.7 \\
80.0 & 40.8 & 120.8 & 39.2 \\
75.0 & 52.6 & 127.6 & 22.4 \\
70.0 & 60.7 & 130.7 & 9.3 \\
65.7 & 65.7 & 131.4 & 0.0 \\
\hline
  \end{tabular}
\end{table}

The spectra obtained for J1940 are not of sufficient resolution to determine a $v\sin i$ for the star. We are, therefore, unable to constrain the values of $i$ and $\beta$ through observations. However, through testing all possible combinations, the best fitting values of $i$ and $\beta$ to equation\,(\ref{eq:OPM}) are $i=31^\circ$ and $\beta=83^\circ$.

Using these values for $i$ and $\beta$, we apply the method of \citet{kurtz92}, based on work by \citet{shibahashi93}, to deconvolve the pulsation into the components of a spherical harmonic series. This technique separates the distorted mode into its pure $\ell=$ 0, 1, 2, \ldots\, spherical harmonic components, allowing us to see the shape of the mode. The results of this deconvolution are shown in Tables\,\ref{tab:har} and \ref{tab:har_comp}. The results show that the mode is a quadrupole mode with a very strong spherically symmetric distortion represented by the radial component. The $\ell=1$ component is small in comparison.

\begin{table}
\setlength\extrarowheight{2pt}
\centering
  \caption{Components of the spherical harmonic series description of the pulsation for $i=31^\circ$ and $\beta=83^\circ$.}
  \label{tab:har}
  \begin{tabular}{lrrr}
    \hline
    $\ell$ & \multicolumn{1}{c}{0} & \multicolumn{1}{c}{1} & \multicolumn{1}{c}{2}  \\
    \hline
    $A_{-2}^{(\ell)}$ (mmag)	&        	&        	&  2.551	\\
    $A_{-1}^{(\ell)}$ (mmag) 	&        	& 1.271  	&  2.410	\\
    $A_0^{(\ell)}$ (mmag)     	&  16.817 	& 0.478  	& -8.146	\\
    $A_{+1}^{(\ell)}$ (mmag)	&        	& 1.066  	& 2.128	\\
    $A_{+2}^{(\ell)}$ (mmag) 	&        	&        	& 1.987	\\
    $\phi$  (rad)                     	& -1.910 	& 2.043  	& -1.775	\\
    \hline
  \end{tabular}
\end{table}

\begin{table}
\centering
  \caption{Comparison between the observed amplitudes and phases to those calculated with the spherical harmonic fit.}
  \label{tab:har_comp}
  \begin{tabular}{lccrr}
    \hline
    ID & 				$A_{\rm obs}$ 	    & $A_{\rm calc}$ 	& \multicolumn{1}{c}{$\phi_{\rm obs}$ }	& \multicolumn{1}{c}{$\phi_{\rm calc}$} \\
    \hline
    $\nu-2\nu_{\rm rot}$ 	& $2.55\pm0.09$ & 2.55			& $  -1.776\pm0.035$ 				& -1.776 \\
    $\nu-1\nu_{\rm rot}$ 	& $1.62\pm0.10$ & 1.62			& $  -2.276\pm0.057$ 				& -2.276 \\
    $\nu$              		& $8.54\pm0.10$ & 8.54 			& $ -2.080\pm0.011$ 				& -2.080 \\ 
    $\nu+1\nu_{\rm rot}$ 	& $2.10\pm0.10$ & 1.45 			& $  -2.276\pm0.044$ 				& -2.239 \\
    $\nu+2\nu_{\rm rot}$ 	& $1.99\pm0.09$ & 1.99 			& $ -3.110\pm0.045$ 				& -1.776 \\
    \hline
  \end{tabular}
\end{table}


\section{Modelling the amplitude and phase variations}
\label{sec:modelling}

The amplitude and phase modulations of roAp stars, which give rise to the rotational sidelobes in the periodogram are explained by the oblique pulsator model, in which the pulsation is axisymmetric with respect to the magnetic axis that is inclined to the rotation axis. The observed pulsation amplitude at a given time is the integral over the amplitude distribution on the visible hemisphere of the star, such that the observed pulsation amplitude and phase modulate as the aspect of the pulsation (and hence magnetic) axis changes as the star rotates. Typically, the pulsation phase changes by about $\pi$-rad at amplitude minima during one rotation period \citep[e.g.][]{kurtz00,holdsworth16}, as a pulsation node crosses the line-of-sight. In contrast, the phase modulations of J1940, as well as HD\,24355 (discussed in \citealt{holdsworth16}), are small and smooth, indicating that the amplitude distribution on the surface deviates considerably from a single spherical harmonic.

The presence of four sidelobe frequencies (at $\pm1\nu_{\rm rot}$ and $\pm2\nu_{\rm rot}$ of the central frequency), coupled with the very large amplitude of the central frequency, indicates that J1940 pulsates in a distorted quadrupole mode with a significant contribution from the spherical-symmetric component. To model such a distorted pulsation, we numerically solve the eigenvalue problem for non-adiabatic linear pulsations under a dipole magnetic field \citep{saio05}, in which the eigenfunction is expanded as a sum of terms proportional to $Y_\ell^0$ with $\ell = 0, 2, 4, \ldots , 38$, including 20 components.

We have searched for models which reproduce the pulsational amplitude and phase modulations of J1940 in the same way as in the case of HD\,24355 \citep{holdsworth16}. For each evolutionary model, assuming a value of $B_{\rm p}$ (the magnetic field strength at poles), we obtain an axisymmetric mode whose frequency is similar to the main frequency of J1940. From the eigenfunction we obtained the amplitude modulation and a set of rotational sidelobe amplitudes. We assume values of the obliquity angle, $\beta$, and of the inclination angle, $i$. We tested many values of $(\beta,i)$ to try and obtain reasonable fits with the relative amplitudes of the rotational sidelobes and amplitude modulation measured in J1940.

Generally, minimum amplitude increases with decreasing $i$, and the local maximum of amplitude at a rotational phase of 0.5 increases with increasing $\beta$ (for $i < 60^\circ$). If we obtain a set of $(\beta, i)$ with which the predicted amplitude modulation approximately reproduces the observed one, we calculate the phase modulation using the parameter set and see whether it is consistent with the observed phase modulation. This process was then repeated for different values of $B_{\rm p}$ to obtain the best fit.

\begin{figure}
   \includegraphics[width=\columnwidth]{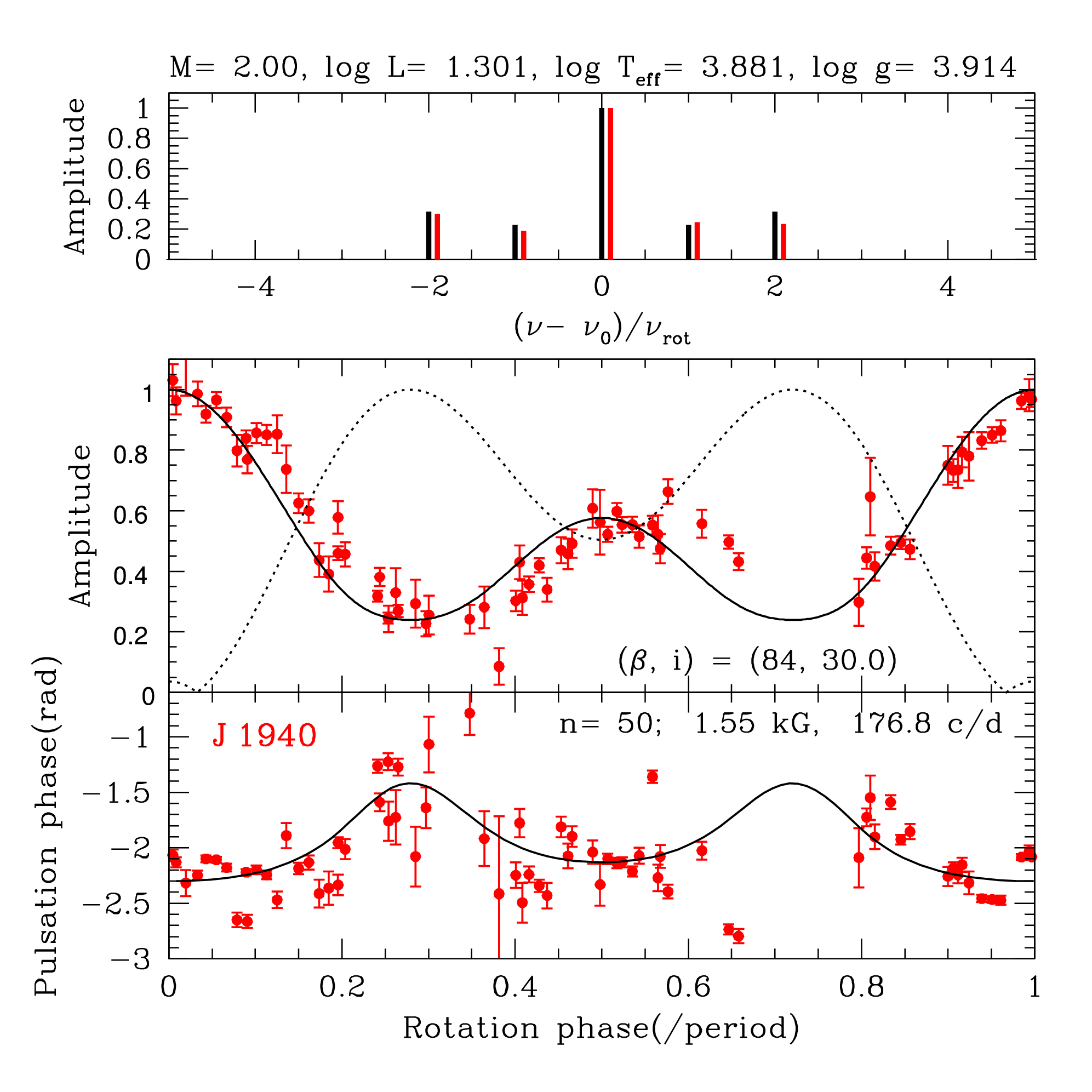}  
\caption{Comparison of amplitude and phase modulations between a model and observational results of J1940. Red and black colours are used for J1940 and the model results, respectively. Top: comparison of the rotational sidelobes, in which the observations are slightly shifted rightward for visibility. Middle: comparison of the amplitude modulations as a function of the rotation phase. The solid line is a theoretical amplitude modulation computed by taking into account the magnetic effects, while the  dotted line indicates the amplitude modulation expected from a pure quadrupole. Bottom: comparison of the modulations of the pulsation phase as a function of the rotation phase; the black line represents the model result, while the red dots with error bars are observational results for J1940.}
\label{fig:modulation}
\end{figure}

Fig.\,\ref{fig:modulation} shows an example of the cases where the theoretical predictions approximately reproduce the observed amplitude and phase modulations of J1940. In most cases, the amplitude modulation can be fitted easily for a certain range of $B_{\rm p}$. However, it is difficult to obtain a theoretical phase modulation similar to the observed one. This is only possible if an appropriate value of $B_{\rm p}$ is chosen for a model whose $T_{\rm eff}$ is within a certain range depending on the modelled mass. In many cases, theoretical phase peaks around rotation phases of 0.3 and 0.7 are less pronounced than those of J1940. 

\begin{figure}
   \includegraphics[width=\columnwidth]{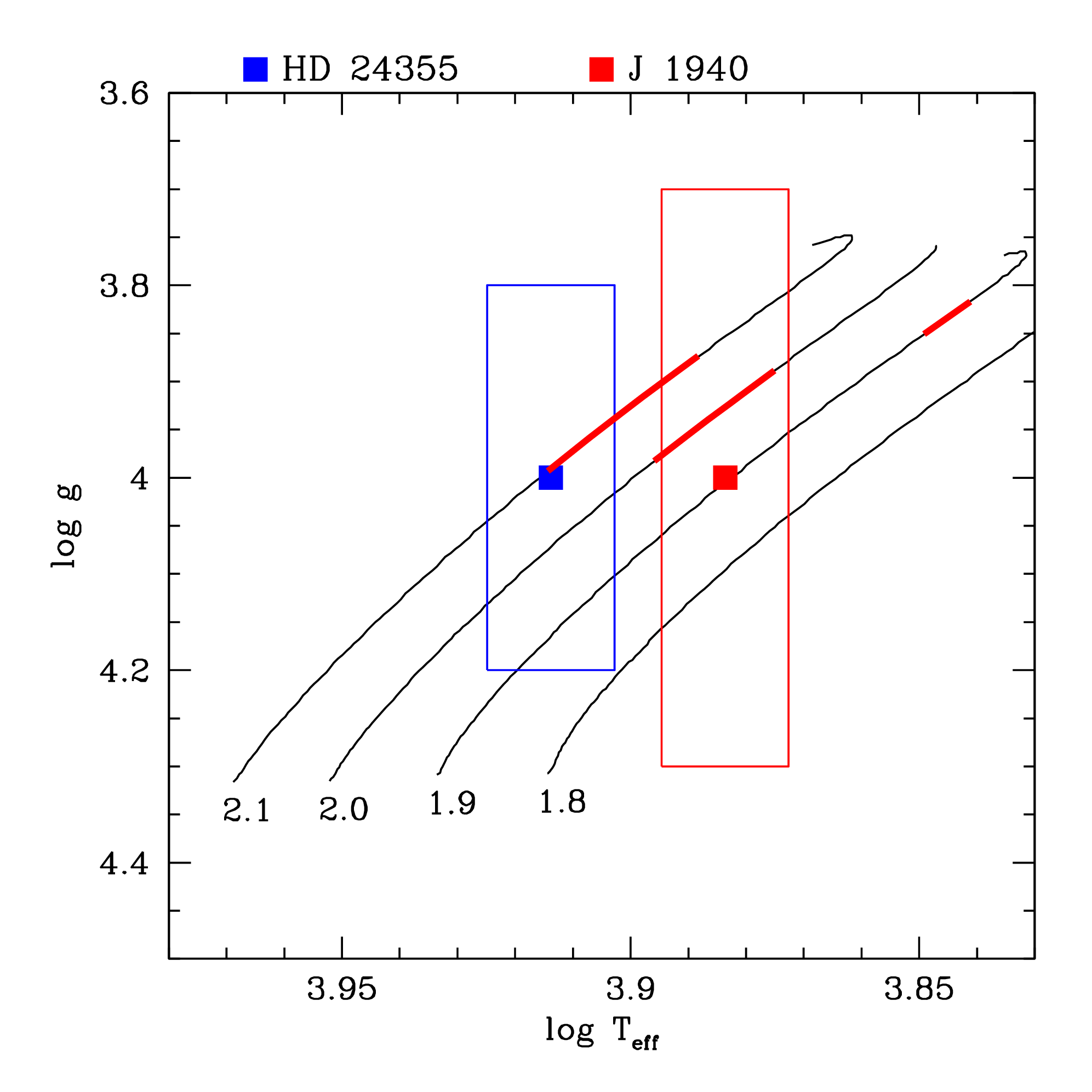}  
   \caption{The $\log T_{\rm eff}-\log g$ diagram showing the approximate positions of J1940 and HD\,24355 with error boxes, and evolutionary tracks of $1.8, 1.9, 2.0,$ and $2.1$\,M$_\odot$ stars. The thick red lines along the evolutionary tracks indicate the positions where theoretical amplitude/phase modulation curves reproduce satisfactorily the observed ones of J1940.}
\label{fig:telgg}
\end{figure}

Fig.\,\ref{fig:telgg} shows ranges (thick red lines) where theoretical phase (and amplitude) modulations are comparable with that shown in Fig.\,\ref{fig:modulation}. To achieve these fits, appropriate values of $B_{\rm p}$ must be chosen. The required $B_{\rm p}$ varies slightly with mass; $1.4\sim1.5$\,kG for $2.1$\,M$_\odot$,  $1.55\sim1.7$\,kG for $2.0$\,M$_\odot$, and $1.8\sim1.9$\,kG for $1.9$\,M$_\odot$. The observed frequencies at around 176\,\cd\, are well above the acoustic cut-off frequencies (116\,\cd) in those models, as in the case of HD\,24355 \citep{holdsworth16}.

The parameters adopted in Fig.\,\ref{fig:modulation} for J1940 yield $\nu L/M = 20.4$, where $\nu$ is the pulsation frequency in mHz, luminosity $L$ and mass $M$ are in solar units, while $\nu L/M=24.7$ for HD\,24355 \citep{holdsworth16}.  The positions in the  $\log T_{\rm eff} - \nu L/M$ plane (figure\,19 of \citealt{holdsworth16}) segregate the roAp stars into some groups. J1940, HD\,24355, and HD\,42659 ($\nu L/M \approx 25$, the star is poorly investigated so far; \citealt{martinez94}) seem  to form a distinct group having high $T_{\rm eff}$ and highest $\nu L/M$, as derived from the models. Interestingly, all are single-mode pulsators. Furthermore, the pulsation mode is classified as a distorted quadrupole mode for both J1940 and HD\,24355. Further observations for the third member HD\,42659 (its binary nature was recently discovered by \citealt{hartmann15}) are highly desirable; it would be interesting to see whether the mode can be also classified as a distorted quadrupole. The presence of the group consisting of highly super-critical single-mode roAp stars might hint that there is yet another excitation mechanism at work in roAp stars.


\section{Summary and Conclusions}

We have presented here the best data, to date, of the rapidly oscillating Ap star J1940. The star was initially discovered to be a roAp star through a survey of A stars in the SuperWASP archive. Amongst the other roAp stars found in that survey, J1940 showed the largest amplitude in the broadband photometry. When considering the relation between pulsation amplitude and observed wavelength for the roAp stars, J1940 was expected to be the largest amplitude roAp star known, when considering standard $B$-band observations for these stars \citep{medupe98}.

The SuperWASP data provided a long time-base of observations to determine the rotation period of the star. Due to the chemical spots on Ap stars, the light curve is modulated with the rotation period of the star. Assuming the rigid rotator model \citep{stibbs50}, we derive a rotation period of the star to be $9.5344\pm0.0012$\,d. 

During the 2014 and 2015 observing seasons, J1940 was subject to single site observations from the South African Astronomical Observatory. Analysis of these data sets suggested that J1940 was a quadrupole pulsator, and confirmed it to have the highest pulsation amplitude amongst the roAp stars, with a peak-to-peak amplitude of 34\,mmag.

As a result of the single site observations, J1940 became only the second roAp star to be the subject of a Whole Earth Telescope campaign, in 2016. Observations over a three week period, from three sites, provided a data set with a duty cycle of 21\,per\,cent. These observations enabled the unambiguous determination of the pulsation frequency and its four rotationally split sidelobes, a result of the quadrupole pulsation in this star.

Further to the quadrupole signature extracted from the data, there is evidence of further rotationally split sidelobes in the light curve. The presence of these sidelobes indicates that J1940 is pulsating in a distorted quadrupole mode.

To achieve the highest frequency resolution analysis, we combined all observations of J1940 into a single light curve, thus providing a frequency resolution of $1.4\times10^{-4}$\,\cd\, ($1.6$\,nHz). In doing so, we attain nearly complete coverage of the rotation cycle of the star, hence allowing us to analyse the pulsation from varying aspects. In doing so we have provided further evidence that J1940 is a distorted pulsator through (i) the fact that the pulsation amplitude does not go to zero at quadrature as a node passes the line-of-sight, and (ii) the pulsation phase does not flip by $\pi$-rad at quadrature, rather it stays almost constant over the rotation cycle of the star.

This phase pattern is similar to other, well studied, quadrupole roAp stars. In Fig.\,\ref{fig:other_quad} we show HD\,24355, KIC\,7582608 and KIC\,10483436 which all exhibit suppressed phase variations over their respective rotation periods which are reminiscent of those seen in J1940. 

To understand this phase behaviour, we modelled J1940 using the method of \citet{saio05}. For a range of model parameters (M, R, $T_{\rm eff}$, $B_{\rm p}$, $i$, $\beta$) we are able to reconstruct the observed amplitude variations well. The modelled amplitude variations deviate significantly from a pure quadrupole mode, confirming that J1940 is pulsating in a distorted quadrupole mode, with a magnetic field strength of about 1.5\,kG.

The pulsation phase variations, however, were somewhat more difficult to model. The peaks in the theoretical phases are lower than those observed (see Fig.\,\ref{fig:modulation}), with the best fitting model being sensitive to both mass and polar magnetic field strength. Finally, the modelling shows that the pulsations in J1940 are well above the theoretical acoustic cut-off frequency. This places J1940 amongst other other roAp stars in the $\log T_{\rm eff}-\nu L/M$ plane. The variations presented here, and the similarities between J1940, HD\,24355 and HD\,42659, may suggest there is another mechanism at work in these stars to excite the pulsations.

We hope to revisit J1940 with observations made by the {\it TESS} mission. With the potential to obtain simultaneous ground-based spectroscopic observations with the space-based photometry, we would be able to answer some of the questions posed in this work. High-precision photometric observations, with a continuous data set for at least 30\,d, would allow us to fully exploit the higher order rotational sidelobes, and to accurately determine any offset between the pulsation and magnetic axes.

\section*{Acknowledgements}

We thank the referee for useful comments and suggestions. DLH acknowledges financial support from the STFC via grant ST/M000877/1. 
BL wishes to thank the Thuthuka grant of the National Research Foundation (NRF) of South Africa. 
This paper uses observations made at the South African Astronomical Observatory (SAAO) and observations obtained with the SARA Observatory 0.9 m telescope at CTIO, which is owned and operated by the Southeastern Association for Research in Astronomy (saraobservatory.org). Thanks go to Alexandre David-Uraz, Rebecca MacInnis, and Trisha Doyle who participated in the SARA observations.
Some of the observations reported in this paper were obtained with the Southern African Large Telescope (SALT) under programmes 2012-2-UKSC-001 and 2015-1-SCI-001. 
The WASP project is funded and maintained by Queen's University Belfast, the Universities of Keele, St. Andrews and Leicester, the Open University, the Isaac Newton Group, the Instituto de Astrofisica Canarias, the South African Astronomical Observatory and by the STFC.

\bibliography{J1940-refs}

\begin{thebibliography}{}
\makeatletter
\relax
\def\mn@urlcharsother{\let\do\@makeother \do\$\do\&\do\#\do\^\do\_\do\%\do\~}
\def\mn@doi{\begingroup\mn@urlcharsother \@ifnextchar [ {\mn@doi@}
  {\mn@doi@[]}}
\def\mn@doi@[#1]#2{\def\@tempa{#1}\ifx\@tempa\@empty \href
  {http://dx.doi.org/#2} {doi:#2}\else \href {http://dx.doi.org/#2} {#1}\fi
  \endgroup}
\def\mn@eprint#1#2{\mn@eprint@#1:#2::\@nil}
\def\mn@eprint@arXiv#1{\href {http://arxiv.org/abs/#1} {{\tt arXiv:#1}}}
\def\mn@eprint@dblp#1{\href {http://dblp.uni-trier.de/rec/bibtex/#1.xml}
  {dblp:#1}}
\def\mn@eprint@#1:#2:#3:#4\@nil{\def\@tempa {#1}\def\@tempb {#2}\def\@tempc
  {#3}\ifx \@tempc \@empty \let \@tempc \@tempb \let \@tempb \@tempa \fi \ifx
  \@tempb \@empty \def\@tempb {arXiv}\fi \@ifundefined
  {mn@eprint@\@tempb}{\@tempb:\@tempc}{\expandafter \expandafter \csname
  mn@eprint@\@tempb\endcsname \expandafter{\@tempc}}}

\bibitem[\protect\citeauthoryear{{Alentiev}, {Kochukhov}, {Ryabchikova},
  {Cunha}, {Tsymbal}  \& {Weiss}}{{Alentiev} et~al.}{2012}]{alentiev12}
{Alentiev} D.,  {Kochukhov} O.,  {Ryabchikova} T.,  {Cunha} M.,  {Tsymbal} V.,
   {Weiss} W.,  2012, \mn@doi [\mnras] {10.1111/j.1745-3933.2011.01211.x},
  \href {http://adsabs.harvard.edu/abs/2012MNRAS.421L..82A} {421, L82}

\bibitem[\protect\citeauthoryear{{Babcock}}{{Babcock}}{1960}]{babcock60}
{Babcock} H.~W.,  1960, \mn@doi [\apj] {10.1086/146960}, \href
  {http://adsabs.harvard.edu/abs/1960ApJ...132..521B} {132, 521}

\bibitem[\protect\citeauthoryear{{Bakos}, {Noyes}, {Kov{\'a}cs}, {Stanek},
  {Sasselov}  \& {Domsa}}{{Bakos} et~al.}{2004}]{bakos04}
{Bakos} G.,  {Noyes} R.~W.,  {Kov{\'a}cs} G.,  {Stanek} K.~Z.,  {Sasselov}
  D.~D.,   {Domsa} I.,  2004, \mn@doi [\pasp] {10.1086/382735}, \href
  {http://adsabs.harvard.edu/abs/2004PASP..116..266B} {116, 266}

\bibitem[\protect\citeauthoryear{{Balmforth}, {Cunha}, {Dolez}, {Gough}  \&
  {Vauclair}}{{Balmforth} et~al.}{2001}]{balmforth01}
{Balmforth} N.~J.,  {Cunha} M.~S.,  {Dolez} N.,  {Gough} D.~O.,   {Vauclair}
  S.,  2001, \mn@doi [\mnras] {10.1046/j.1365-8711.2001.04182.x}, \href
  {http://ukads.nottingham.ac.uk/abs/2001MNRAS.323..362B} {323, 362}

\bibitem[\protect\citeauthoryear{{Balona} et~al.,}{{Balona}
  et~al.}{2011a}]{balona11a}
{Balona} L.~A.,  et~al., 2011a, \mn@doi [\mnras]
  {10.1111/j.1365-2966.2010.17461.x}, \href
  {http://adsabs.harvard.edu/abs/2011MNRAS.410..517B} {410, 517}

\bibitem[\protect\citeauthoryear{{Balona} et~al.,}{{Balona}
  et~al.}{2011b}]{balona11b}
{Balona} L.~A.,  et~al., 2011b, \mn@doi [\mnras]
  {10.1111/j.1365-2966.2011.18334.x}, \href
  {http://adsabs.harvard.edu/abs/2011MNRAS.413.2651B} {413, 2651}

\bibitem[\protect\citeauthoryear{{Bertelli}, {Girardi}, {Marigo}  \&
  {Nasi}}{{Bertelli} et~al.}{2008}]{bertelli08}
{Bertelli} G.,  {Girardi} L.,  {Marigo} P.,   {Nasi} E.,  2008, \mn@doi [\aap]
  {10.1051/0004-6361:20079165}, \href
  {http://adsabs.harvard.edu/abs/2008A%26A...484..815B} {484, 815}

\bibitem[\protect\citeauthoryear{{Bigot} \& {Dziembowski}}{{Bigot} \&
  {Dziembowski}}{2002}]{bigot02}
{Bigot} L.,  {Dziembowski} W.~A.,  2002, \mn@doi [\aap]
  {10.1051/0004-6361:20020824}, \href
  {http://adsabs.harvard.edu/abs/2002A%26A...391..235B} {391, 235}

\bibitem[\protect\citeauthoryear{{Bigot} \& {Kurtz}}{{Bigot} \&
  {Kurtz}}{2011}]{bigot11}
{Bigot} L.,  {Kurtz} D.~W.,  2011, \mn@doi [\aap]
  {10.1051/0004-6361/201116981}, \href
  {http://adsabs.harvard.edu/abs/2011A%26A...536A..73B} {536, A73}

\bibitem[\protect\citeauthoryear{{Coppejans} et~al.,}{{Coppejans}
  et~al.}{2013}]{coppejans13}
{Coppejans} R.,  et~al., 2013, \mn@doi [\pasp] {10.1086/672156}, \href
  {http://adsabs.harvard.edu/abs/2013PASP..125..976C} {125, 976}

\bibitem[\protect\citeauthoryear{{Cowley}, {Hubrig}, {Ryabchikova}, {Mathys},
  {Piskunov}  \& {Mittermayer}}{{Cowley} et~al.}{2001}]{cowley01}
{Cowley} C.~R.,  {Hubrig} S.,  {Ryabchikova} T.~A.,  {Mathys} G.,  {Piskunov}
  N.,   {Mittermayer} P.,  2001, \mn@doi [\aap] {10.1051/0004-6361:20000539},
  \href {http://adsabs.harvard.edu/abs/2001A%26A...367..939C} {367, 939}

\bibitem[\protect\citeauthoryear{{Cunha}, {Alentiev}, {Brand{\~a}o}  \&
  {Perraut}}{{Cunha} et~al.}{2013}]{cunha13}
{Cunha} M.~S.,  {Alentiev} D.,  {Brand{\~a}o} I.~M.,   {Perraut} K.,  2013,
  \mn@doi [\mnras] {10.1093/mnras/stt1679}, \href
  {http://adsabs.harvard.edu/abs/2013MNRAS.436.1639C} {436, 1639}

\bibitem[\protect\citeauthoryear{{Dalessio}}{{Dalessio}}{2010}]{dalessio10}
{Dalessio} J.,  2010, in American Astronomical Society Meeting Abstracts \#215.
  p.~462

\bibitem[\protect\citeauthoryear{{Dorokhova} \& {Dorokhov}}{{Dorokhova} \&
  {Dorokhov}}{2005}]{dorokhova05}
{Dorokhova} T.,  {Dorokhov} N.,  2005, \mn@doi [Journal of Astrophysics and
  Astronomy] {10.1007/BF02702330}, \href
  {http://adsabs.harvard.edu/abs/2005JApA...26..223D} {26, 223}

\bibitem[\protect\citeauthoryear{{Dziembowski} \& {Goode}}{{Dziembowski} \&
  {Goode}}{1985}]{dg85}
{Dziembowski} W.,  {Goode} P.~R.,  1985, \mn@doi [\apjl] {10.1086/184542},
  \href {http://ukads.nottingham.ac.uk/abs/1985ApJ...296L..27D} {296, L27}

\bibitem[\protect\citeauthoryear{{Elkin}, {Kurtz}, {Mathys}  \&
  {Freyhammer}}{{Elkin} et~al.}{2010}]{elkin10}
{Elkin} V.~G.,  {Kurtz} D.~W.,  {Mathys} G.,   {Freyhammer} L.~M.,  2010,
  \mn@doi [\mnras] {10.1111/j.1745-3933.2010.00844.x}, \href
  {http://adsabs.harvard.edu/abs/2010MNRAS.404L.104E} {404, L104}

\bibitem[\protect\citeauthoryear{{Elkin}, {Kurtz}, {Worters}, {Mathys},
  {Smalley}, {van Wyk}  \& {Smith}}{{Elkin} et~al.}{2011}]{elkin11}
{Elkin} V.~G.,  {Kurtz} D.~W.,  {Worters} H.~L.,  {Mathys} G.,  {Smalley} B.,
  {van Wyk} F.,   {Smith} A.~M.~S.,  2011, \mn@doi [\mnras]
  {10.1111/j.1365-2966.2010.17747.x}, \href
  {http://adsabs.harvard.edu/abs/2011MNRAS.411..978E} {411, 978}

\bibitem[\protect\citeauthoryear{{Handler} \& {Paunzen}}{{Handler} \&
  {Paunzen}}{1999}]{handler99}
{Handler} G.,  {Paunzen} E.,  1999, \mn@doi [\aaps] {10.1051/aas:1999159},
  \href {http://adsabs.harvard.edu/abs/1999A%26AS..135...57H} {135, 57}

\bibitem[\protect\citeauthoryear{{Hartman}, {Bakos}, {Noyes}, {Sip{\H o}cz},
  {Kov{\'a}cs}, {Mazeh}, {Shporer}  \& {P{\'a}l}}{{Hartman}
  et~al.}{2011}]{hartman11}
{Hartman} J.~D.,  {Bakos} G.~{\'A}.,  {Noyes} R.~W.,  {Sip{\H o}cz} B.,
  {Kov{\'a}cs} G.,  {Mazeh} T.,  {Shporer} A.,   {P{\'a}l} A.,  2011, \mn@doi
  [\aj] {10.1088/0004-6256/141/5/166}, \href
  {http://adsabs.harvard.edu/abs/2011AJ....141..166H} {141, 166}

\bibitem[\protect\citeauthoryear{{Hartmann} \& {Hatzes}}{{Hartmann} \&
  {Hatzes}}{2015}]{hartmann15}
{Hartmann} M.,  {Hatzes} A.~P.,  2015, \mn@doi [\aap]
  {10.1051/0004-6361/201425320}, \href
  {http://adsabs.harvard.edu/abs/2015A%26A...582A..84H} {582, A84}

\bibitem[\protect\citeauthoryear{{Hatzes} \& {Mkrtichian}}{{Hatzes} \&
  {Mkrtichian}}{2004}]{hatzes04}
{Hatzes} A.~P.,  {Mkrtichian} D.~E.,  2004, \mn@doi [\mnras]
  {10.1111/j.1365-2966.2004.07816.x}, \href
  {http://ukads.nottingham.ac.uk/abs/2004MNRAS.351..663H} {351, 663}

\bibitem[\protect\citeauthoryear{{Holdsworth}}{{Holdsworth}}{2015}]{holdsworth15}
{Holdsworth} D.~L.,  2015, PhD thesis, Keele University, UK

\bibitem[\protect\citeauthoryear{{Holdsworth}}{{Holdsworth}}{2016}]{holdsworth16b}
{Holdsworth} D.~L.,  2016, Information Bulletin on Variable Stars, \href
  {http://adsabs.harvard.edu/abs/2016IBVS.6185....1P} {6185}

\bibitem[\protect\citeauthoryear{{Holdsworth} et~al.,}{{Holdsworth}
  et~al.}{2014a}]{holdsworth14a}
{Holdsworth} D.~L.,  et~al., 2014a, \mn@doi [\mnras] {10.1093/mnras/stu094},
  \href {http://adsabs.harvard.edu/abs/2014MNRAS.439.2078H} {439, 2078}

\bibitem[\protect\citeauthoryear{{Holdsworth}, {Smalley}, {Kurtz},
  {Southworth}, {Cunha}  \& {Clubb}}{{Holdsworth}
  et~al.}{2014b}]{holdsworth14b}
{Holdsworth} D.~L.,  {Smalley} B.,  {Kurtz} D.~W.,  {Southworth} J.,  {Cunha}
  M.~S.,   {Clubb} K.~I.,  2014b, \mn@doi [\mnras] {10.1093/mnras/stu1303},
  \href {http://adsabs.harvard.edu/abs/2014MNRAS.443.2049H} {443, 2049}

\bibitem[\protect\citeauthoryear{{Holdsworth}, {Kurtz}, {Smalley}, {Saio},
  {Handler}, {Murphy}  \& {Lehmann}}{{Holdsworth} et~al.}{2016}]{holdsworth16}
{Holdsworth} D.~L.,  {Kurtz} D.~W.,  {Smalley} B.,  {Saio} H.,  {Handler} G.,
  {Murphy} S.~J.,   {Lehmann} H.,  2016, \mn@doi [\mnras]
  {10.1093/mnras/stw1711}, \href
  {http://adsabs.harvard.edu/abs/2016MNRAS.462..876H} {462, 876}

\bibitem[\protect\citeauthoryear{{Holdsworth}, {{\O}stensen}, {Smalley}  \&
  {Telting}}{{Holdsworth} et~al.}{2017}]{holdsworth17a}
{Holdsworth} D.~L.,  {{\O}stensen} R.~H.,  {Smalley} B.,   {Telting} J.~H.,
  2017, \mn@doi [\mnras] {10.1093/mnras/stx077}, \href
  {http://adsabs.harvard.edu/abs/2017MNRAS.tmp...86H} {466, 5020}

\bibitem[\protect\citeauthoryear{{Howell} et~al.,}{{Howell}
  et~al.}{2014}]{howell14}
{Howell} S.~B.,  et~al., 2014, \mn@doi [\pasp] {10.1086/676406}, \href
  {http://adsabs.harvard.edu/abs/2014PASP..126..398H} {126, 398}

\bibitem[\protect\citeauthoryear{{Joshi} et~al.,}{{Joshi}
  et~al.}{2016}]{joshi16}
{Joshi} S.,  et~al., 2016, \mn@doi [\aap] {10.1051/0004-6361/201527242}, \href
  {http://adsabs.harvard.edu/abs/2016A%26A...590A.116J} {590, A116}

\bibitem[\protect\citeauthoryear{{Kobulnicky}, {Nordsieck}, {Burgh}, {Smith},
  {Percival}, {Williams}  \& {O'Donoghue}}{{Kobulnicky}
  et~al.}{2003}]{kobulnicky03}
{Kobulnicky} H.~A.,  {Nordsieck} K.~H.,  {Burgh} E.~B.,  {Smith} M.~P.,
  {Percival} J.~W.,  {Williams} T.~B.,   {O'Donoghue} D.,  2003, in {Iye} M.,
  {Moorwood} A.~F.~M.,  eds,  \procspie Vol. 4841, Instrument Design and
  Performance for Optical/Infrared Ground-based Telescopes. pp 1634--1644,
  \mn@doi{10.1117/12.460315}

\bibitem[\protect\citeauthoryear{{Kochukhov} \& {Ryabchikova}}{{Kochukhov} \&
  {Ryabchikova}}{2001}]{koch01}
{Kochukhov} O.,  {Ryabchikova} T.,  2001, \mn@doi [\aap]
  {10.1051/0004-6361:20010726}, \href
  {http://ukads.nottingham.ac.uk/abs/2001A%26A...374..615K} {374, 615}

\bibitem[\protect\citeauthoryear{{Kochukhov}, {Alentiev}, {Ryabchikova},
  {Boyko}, {Cunha}, {Tsymbal}  \& {Weiss}}{{Kochukhov}
  et~al.}{2013}]{kochukhov13}
{Kochukhov} O.,  {Alentiev} D.,  {Ryabchikova} T.,  {Boyko} S.,  {Cunha} M.,
  {Tsymbal} V.,   {Weiss} W.,  2013, \mn@doi [\mnras] {10.1093/mnras/stt377},
  \href {http://adsabs.harvard.edu/abs/2013MNRAS.431.2808K} {431, 2808}

\bibitem[\protect\citeauthoryear{{Kochukhov} et~al.,}{{Kochukhov}
  et~al.}{2015}]{kochukhov15}
{Kochukhov} O.,  et~al., 2015, \mn@doi [\aap] {10.1051/0004-6361/201425065},
  \href {http://adsabs.harvard.edu/abs/2015A%26A...574A..79K} {574, A79}

\bibitem[\protect\citeauthoryear{{Kurtz}}{{Kurtz}}{1982}]{kurtz82}
{Kurtz} D.~W.,  1982, \mnras, \href
  {http://adsabs.harvard.edu/abs/1982MNRAS.200..807K} {200, 807}

\bibitem[\protect\citeauthoryear{{Kurtz}}{{Kurtz}}{1992}]{kurtz92}
{Kurtz} D.~W.,  1992, \mn@doi [\mnras] {10.1093/mnras/259.4.701}, \href
  {http://adsabs.harvard.edu/abs/1992MNRAS.259..701K} {259, 701}

\bibitem[\protect\citeauthoryear{{Kurtz} \& {Martinez}}{{Kurtz} \&
  {Martinez}}{2000}]{kurtz00}
{Kurtz} D.~W.,  {Martinez} P.,  2000, Baltic Astronomy, \href
  {http://adsabs.harvard.edu/abs/2000BaltA...9..253K} {9, 253}

\bibitem[\protect\citeauthoryear{{Kurtz}, {Shibahashi}  \& {Goode}}{{Kurtz}
  et~al.}{1990}]{kurtz90}
{Kurtz} D.~W.,  {Shibahashi} H.,   {Goode} P.~R.,  1990, \mnras, \href
  {http://adsabs.harvard.edu/abs/1990MNRAS.247..558K} {247, 558}

\bibitem[\protect\citeauthoryear{{Kurtz}, {Martinez}, {van Wyk}, {Marang}  \&
  {Roberts}}{{Kurtz} et~al.}{1994}]{kurtz94}
{Kurtz} D.~W.,  {Martinez} P.,  {van Wyk} F.,  {Marang} F.,   {Roberts} G.,
  1994, \mn@doi [\mnras] {10.1093/mnras/268.3.641}, \href
  {http://adsabs.harvard.edu/abs/1994MNRAS.268..641K} {268, 641}

\bibitem[\protect\citeauthoryear{{Kurtz}, {van Wyk}, {Roberts}, {Marang},
  {Handler}, {Medupe}  \& {Kilkenny}}{{Kurtz} et~al.}{1997}]{kurtz97}
{Kurtz} D.~W.,  {van Wyk} F.,  {Roberts} G.,  {Marang} F.,  {Handler} G.,
  {Medupe} R.,   {Kilkenny} D.,  1997, \mn@doi [\mnras]
  {10.1093/mnras/287.1.69}, \href
  {http://adsabs.harvard.edu/abs/1997MNRAS.287...69K} {287, 69}

\bibitem[\protect\citeauthoryear{{Kurtz} et~al.,}{{Kurtz}
  et~al.}{2005}]{kurtz05}
{Kurtz} D.~W.,  et~al., 2005, \mn@doi [\mnras]
  {10.1111/j.1365-2966.2005.08807.x}, \href
  {http://adsabs.harvard.edu/abs/2005MNRAS.358..651K} {358, 651}

\bibitem[\protect\citeauthoryear{{Kurtz} et~al.,}{{Kurtz}
  et~al.}{2011}]{kurtz11}
{Kurtz} D.~W.,  et~al., 2011, \mn@doi [\mnras]
  {10.1111/j.1365-2966.2011.18572.x}, \href
  {http://adsabs.harvard.edu/abs/2011MNRAS.414.2550K} {414, 2550}

\bibitem[\protect\citeauthoryear{{Lenz} \& {Breger}}{{Lenz} \&
  {Breger}}{2005}]{lenz05}
{Lenz} P.,  {Breger} M.,  2005, \mn@doi [Commun. Asteroseismol.]
  {10.1553/cia146s53}, \href
  {http://adsabs.harvard.edu/abs/2005CoAst.146...53L} {146, 53}

\bibitem[\protect\citeauthoryear{{L{\"u}ftinger}, {Kochukhov}, {Ryabchikova},
  {Piskunov}, {Weiss}  \& {Ilyin}}{{L{\"u}ftinger}
  et~al.}{2010a}]{lueftinger10}
{L{\"u}ftinger} T.,  {Kochukhov} O.,  {Ryabchikova} T.,  {Piskunov} N.,
  {Weiss} W.~W.,   {Ilyin} I.,  2010a, \mn@doi [\aap]
  {10.1051/0004-6361/200811545}, \href
  {http://adsabs.harvard.edu/abs/2010A%26A...509A..71L} {509, A71}

\bibitem[\protect\citeauthoryear{{L{\"u}ftinger}, {Kochukhov}, {Ryabchikova},
  {Piskunov}, {Weiss}  \& {Ilyin}}{{L{\"u}ftinger} et~al.}{2010b}]{luftinger10}
{L{\"u}ftinger} T.,  {Kochukhov} O.,  {Ryabchikova} T.,  {Piskunov} N.,
  {Weiss} W.~W.,   {Ilyin} I.,  2010b, \mn@doi [\aap]
  {10.1051/0004-6361/200811545}, \href
  {http://adsabs.harvard.edu/abs/2010A%26A...509A..71L} {509, A71}

\bibitem[\protect\citeauthoryear{{Martinez} \& {Kurtz}}{{Martinez} \&
  {Kurtz}}{1994}]{martinez94}
{Martinez} P.,  {Kurtz} D.~W.,  1994, \mnras, \href
  {http://adsabs.harvard.edu/abs/1994MNRAS.271..118M} {271, 118}

\bibitem[\protect\citeauthoryear{{Martinez}, {Kurtz}  \&
  {Kauffmann}}{{Martinez} et~al.}{1991}]{martinez91}
{Martinez} P.,  {Kurtz} D.~W.,   {Kauffmann} G.~M.,  1991, \mnras, \href
  {http://adsabs.harvard.edu/abs/1991MNRAS.250..666M} {250, 666}

\bibitem[\protect\citeauthoryear{{Mathys}}{{Mathys}}{1985}]{mathys85}
{Mathys} G.,  1985, \aap, \href
  {http://adsabs.harvard.edu/abs/1985A%26A...151..315M} {151, 315}

\bibitem[\protect\citeauthoryear{{Mathys}}{{Mathys}}{2016}]{mathys17}
{Mathys} G.,  2016, preprint, \href
  {http://adsabs.harvard.edu/abs/2016arXiv161203632M} {} (\mn@eprint {arXiv}
  {1612.03632})

\bibitem[\protect\citeauthoryear{{Medupe} \& {Kurtz}}{{Medupe} \&
  {Kurtz}}{1998}]{medupe98}
{Medupe} R.,  {Kurtz} D.~W.,  1998, \mn@doi [\mnras]
  {10.1046/j.1365-8711.1998.01772.x}, \href
  {http://adsabs.harvard.edu/abs/1998MNRAS.299..371M} {299, 371}

\bibitem[\protect\citeauthoryear{{Mkrtichian}, {Hatzes}, {Saio}  \&
  {Shobbrook}}{{Mkrtichian} et~al.}{2008}]{mkr08}
{Mkrtichian} D.~E.,  {Hatzes} A.~P.,  {Saio} H.,   {Shobbrook} R.~R.,  2008,
  \mn@doi [\aap] {10.1051/0004-6361:200809890}, \href
  {http://ukads.nottingham.ac.uk/abs/2008A%26A...490.1109M} {490, 1109}

\bibitem[\protect\citeauthoryear{{Monet} et~al.,}{{Monet}
  et~al.}{2003}]{monet03}
{Monet} D.~G.,  et~al., 2003, \mn@doi [\aj] {10.1086/345888}, \href
  {http://adsabs.harvard.edu/abs/2003AJ....125..984M} {125, 984}

\bibitem[\protect\citeauthoryear{{Montgomery} \& {O'Donoghue}}{{Montgomery} \&
  {O'Donoghue}}{1999}]{montgomery99}
{Montgomery} M.~H.,  {O'Donoghue} D.,  1999, Delta Scuti Star Newsletter, \href
  {http://adsabs.harvard.edu/abs/1999DSSN...13...28M} {13, 28}

\bibitem[\protect\citeauthoryear{{Murphy}, {Shibahashi}  \& {Kurtz}}{{Murphy}
  et~al.}{2013}]{murphy13}
{Murphy} S.~J.,  {Shibahashi} H.,   {Kurtz} D.~W.,  2013, \mn@doi [\mnras]
  {10.1093/mnras/stt105}, \href
  {http://adsabs.harvard.edu/abs/2013MNRAS.430.2986M} {430, 2986}

\bibitem[\protect\citeauthoryear{{Nather}, {Winget}, {Clemens}, {Hansen}  \&
  {Hine}}{{Nather} et~al.}{1990}]{nather90}
{Nather} R.~E.,  {Winget} D.~E.,  {Clemens} J.~C.,  {Hansen} C.~J.,   {Hine}
  B.~P.,  1990, \mn@doi [\apj] {10.1086/169196}, \href
  {http://adsabs.harvard.edu/abs/1990ApJ...361..309N} {361, 309}

\bibitem[\protect\citeauthoryear{{Paunzen}, {Netopil}, {Rode-Paunzen},
  {Handler}  \& {Bo{\v z}i{\'c}}}{{Paunzen} et~al.}{2015}]{paunzen15}
{Paunzen} E.,  {Netopil} M.,  {Rode-Paunzen} M.,  {Handler} G.,   {Bo{\v
  z}i{\'c}} H.,  2015, \mn@doi [\aap] {10.1051/0004-6361/201425281}, \href
  {http://adsabs.harvard.edu/abs/2015A%26A...575A..24P} {575, A24}

\bibitem[\protect\citeauthoryear{{Pepper} et~al.,}{{Pepper}
  et~al.}{2007}]{pepper07}
{Pepper} J.,  et~al., 2007, \mn@doi [\pasp] {10.1086/521836}, \href
  {http://adsabs.harvard.edu/abs/2007PASP..119..923P} {119, 923}

\bibitem[\protect\citeauthoryear{{Pepper}, {Stanek}, {Pogge}, {Latham},
  {DePoy}, {Siverd}, {Poindexter}  \& {Sivakoff}}{{Pepper}
  et~al.}{2008}]{pepper08}
{Pepper} J.,  {Stanek} K.~Z.,  {Pogge} R.~W.,  {Latham} D.~W.,  {DePoy} D.~L.,
  {Siverd} R.,  {Poindexter} S.,   {Sivakoff} G.~R.,  2008, \mn@doi [\aj]
  {10.1088/0004-6256/135/3/907}, \href
  {http://adsabs.harvard.edu/abs/2008AJ....135..907P} {135, 907}

\bibitem[\protect\citeauthoryear{{Pojmanski}}{{Pojmanski}}{1997}]{pojmanski97}
{Pojmanski} G.,  1997, \actaa, \href
  {http://adsabs.harvard.edu/abs/1997AcA....47..467P} {47, 467}

\bibitem[\protect\citeauthoryear{{Pollacco} et~al.,}{{Pollacco}
  et~al.}{2006}]{pollacco06}
{Pollacco} D.~L.,  et~al., 2006, \mn@doi [\pasp] {10.1086/508556}, \href
  {http://adsabs.harvard.edu/abs/2006PASP..118.1407P} {118, 1407}

\bibitem[\protect\citeauthoryear{{Provencal} et~al.,}{{Provencal}
  et~al.}{2009}]{provencal09}
{Provencal} J.~L.,  et~al., 2009, \mn@doi [\apj] {10.1088/0004-637X/693/1/564},
  \href {http://adsabs.harvard.edu/abs/2009ApJ...693..564P} {693, 564}

\bibitem[\protect\citeauthoryear{{Provencal} et~al.,}{{Provencal}
  et~al.}{2012}]{provencal12}
{Provencal} J.~L.,  et~al., 2012, \mn@doi [\apj] {10.1088/0004-637X/751/2/91},
  \href {http://adsabs.harvard.edu/abs/2012ApJ...751...91P} {751, 91}

\bibitem[\protect\citeauthoryear{{Ricker} et~al.,}{{Ricker}
  et~al.}{2015}]{ricker15}
{Ricker} G.~R.,  et~al., 2015, \mn@doi [Journal of Astronomical Telescopes,
  Instruments, and Systems] {10.1117/1.JATIS.1.1.014003}, \href
  {http://adsabs.harvard.edu/abs/2015JATIS...1a4003R} {1, 014003}

\bibitem[\protect\citeauthoryear{{Ryabchikova}, {Nesvacil}, {Weiss},
  {Kochukhov}  \& {St{\"u}tz}}{{Ryabchikova} et~al.}{2004}]{ryabchikova04}
{Ryabchikova} T.,  {Nesvacil} N.,  {Weiss} W.~W.,  {Kochukhov} O.,
  {St{\"u}tz} C.,  2004, \mn@doi [\aap] {10.1051/0004-6361:20041012}, \href
  {http://adsabs.harvard.edu/abs/2004A%26A...423..705R} {423, 705}

\bibitem[\protect\citeauthoryear{{Saio}}{{Saio}}{2005}]{saio05}
{Saio} H.,  2005, \mn@doi [\mnras] {10.1111/j.1365-2966.2005.09091.x}, \href
  {http://ukads.nottingham.ac.uk/abs/2005MNRAS.360.1022S} {360, 1022}

\bibitem[\protect\citeauthoryear{{Savanov}, {Malanushenko}  \&
  {Ryabchikova}}{{Savanov} et~al.}{1999}]{savanov99}
{Savanov} I.~S.,  {Malanushenko} V.~P.,   {Ryabchikova} T.~A.,  1999, Astronomy
  Letters, \href {http://ukads.nottingham.ac.uk/abs/1999AstL...25..802S} {25,
  802}

\bibitem[\protect\citeauthoryear{{Shibahashi} \& {Saio}}{{Shibahashi} \&
  {Saio}}{1985a}]{ss85a}
{Shibahashi} H.,  {Saio} H.,  1985a, \pasj, \href
  {http://ukads.nottingham.ac.uk/abs/1985PASJ...37..245S} {37, 245}

\bibitem[\protect\citeauthoryear{{Shibahashi} \& {Saio}}{{Shibahashi} \&
  {Saio}}{1985b}]{ss85b}
{Shibahashi} H.,  {Saio} H.,  1985b, \pasj, \href
  {http://ukads.nottingham.ac.uk/abs/1985PASJ...37..601S} {37, 601}

\bibitem[\protect\citeauthoryear{{Shibahashi} \& {Takata}}{{Shibahashi} \&
  {Takata}}{1993}]{shibahashi93}
{Shibahashi} H.,  {Takata} M.,  1993, \pasj, \href
  {http://ukads.nottingham.ac.uk/abs/1993PASJ...45..617S} {45, 617}

\bibitem[\protect\citeauthoryear{{Smalley} et~al.,}{{Smalley}
  et~al.}{2015}]{smalley15}
{Smalley} B.,  et~al., 2015, \mn@doi [\mnras] {10.1093/mnras/stv1515}, \href
  {http://adsabs.harvard.edu/abs/2015MNRAS.452.3334S} {452, 3334}

\bibitem[\protect\citeauthoryear{{Smalley} et~al.,}{{Smalley}
  et~al.}{2017}]{smalley17}
{Smalley} B.,  et~al., 2017, \mn@doi [\mnras] {10.1093/mnras/stw2903}, \href
  {http://adsabs.harvard.edu/abs/2017MNRAS.465.2662S} {465, 2662}

\bibitem[\protect\citeauthoryear{{Smith} et~al.,}{{Smith}
  et~al.}{2006}]{smith06}
{Smith} A.~M.~S.,  et~al., 2006, \mn@doi [\mnras]
  {10.1111/j.1365-2966.2006.11095.x}, \href
  {http://adsabs.harvard.edu/abs/2006MNRAS.373.1151S} {373, 1151}

\bibitem[\protect\citeauthoryear{{Stibbs}}{{Stibbs}}{1950}]{stibbs50}
{Stibbs} D.~W.~N.,  1950, \mnras, \href
  {http://adsabs.harvard.edu/abs/1950MNRAS.110..395S} {110, 395}

\bibitem[\protect\citeauthoryear{{Stoehr} et~al.,}{{Stoehr}
  et~al.}{2008}]{stoehr08}
{Stoehr} F.,  et~al., 2008, in {Argyle} R.~W.,  {Bunclark} P.~S.,   {Lewis}
  J.~R.,  eds,  Astronomical Society of the Pacific Conference Series Vol. 394,
  Astronomical Data Analysis Software and Systems XVII. p.~505

\bibitem[\protect\citeauthoryear{{Takata} \& {Shibahashi}}{{Takata} \&
  {Shibahashi}}{1994}]{ts94}
{Takata} M.,  {Shibahashi} H.,  1994, \pasj, \href
  {http://ukads.nottingham.ac.uk/abs/1994PASJ...46..301T} {46, 301}

\bibitem[\protect\citeauthoryear{{Takata} \& {Shibahashi}}{{Takata} \&
  {Shibahashi}}{1995}]{ts95}
{Takata} M.,  {Shibahashi} H.,  1995, \pasj, \href
  {http://ukads.nottingham.ac.uk/abs/1995PASJ...47..219T} {47, 219}

\bibitem[\protect\citeauthoryear{{Thompson} \& {Mullally}}{{Thompson} \&
  {Mullally}}{2009}]{thompson09}
{Thompson} S.~E.,  {Mullally} F.,  2009, in Journal of Physics Conference
  Series. p. 012081, \mn@doi{10.1088/1742-6596/172/1/012081}

\bibitem[\protect\citeauthoryear{{Udalski}, {Szymanski}, {Kaluzny}, {Kubiak}
  \& {Mateo}}{{Udalski} et~al.}{1992}]{udalski92}
{Udalski} A.,  {Szymanski} M.,  {Kaluzny} J.,  {Kubiak} M.,   {Mateo} M.,
  1992, \actaa, \href {http://adsabs.harvard.edu/abs/1992AcA....42..253U} {42,
  253}

\bibitem[\protect\citeauthoryear{{Ulaczyk} et~al.,}{{Ulaczyk}
  et~al.}{2013}]{ulaczyk13}
{Ulaczyk} K.,  et~al., 2013, \actaa, \href
  {http://adsabs.harvard.edu/abs/2013AcA....63..159U} {63, 159}

\makeatother
\end{thebibliography}

\label{lastpage}
\end{document}